\documentclass[a4paper,11pt]{article}
\pdfoutput=1 \usepackage{jheppub_2} 
\usepackage{booktabs}
\usepackage[T1]{fontenc} 
\usepackage{array}
\usepackage{graphicx, multirow,soul,url,amsmath,amsfonts,amssymb,mathrsfs,amsfonts}
\usepackage[all]{hypcap}
\usepackage{url}
\usepackage{bbm}
\usepackage{changepage}
\usepackage[section]{placeins}
\usepackage{cancel,ulem}
\usepackage{sidecap}
\usepackage{gensymb}
\usepackage{graphicx}
\usepackage{slashed}
\usepackage[dvipsnames]{xcolor}
\usepackage{makecell}
\usepackage{multicol, blindtext}
\usepackage{multirow}
\usepackage{setspace}
\usepackage[section]{placeins}
\usepackage[version=3]{mhchem}
\usepackage{caption}
\expandafter\def\csname ver@subfig.sty\endcsname{}
\usepackage{svg}
\usepackage{lipsum}
\usepackage{hyperref}
\usepackage{footnote}
\usepackage{bbold}
\usepackage{amsbsy}
\usepackage[export]{adjustbox}
\usepackage{units}

\newcommand{\LCDM}{\Lambda\text{CDM}}
\usepackage{float}
\usepackage{pifont}
\restylefloat{table}
\usepackage{lipsum}

\def\thefootnote{\fnsymbol{footnote}}
\allowdisplaybreaks

\newcommand{\be}{\begin{equation}}
\newcommand{\ee}{\end{equation}}
\newcommand{\bea}{\begin{eqnarray}}
\newcommand{\eea}{\end{eqnarray}}
\newcommand{\ba}{\begin{aligned}}
\newcommand{\ea}{\end{aligned}}

\newcommand{\bq}{\begin{eqnarray}}
\newcommand{\nq}{\end{eqnarray}}

\usepackage{subfig}

\title{\bf Impact of cosmic evolution on the gravitational wave spectrum from cosmic strings}

\author[a,b,c]{Xi-He Hu,}
\affiliation[a]{School of Fundamental Physics and Mathematical Sciences, Hangzhou Institute for Advanced Study, UCAS, Hangzhou 310024, China}
\affiliation[b]{Institute of Theoretical Physics, Chinese Academy of Sciences, Beijing 100190, China}
\affiliation[c]{University of Chinese Academy of Sciences, Beijing 100049, China}
\author[d]{Danny Marfatia}
\affiliation[d]{Department of Physics and Astronomy, University of Hawaii at Manoa, Honolulu, HI 96822, USA}
\author[a]{and Ye-Ling Zhou}
\emailAdd{huxihe23@mails.ucas.ac.cn}
\emailAdd{dmarf8@hawaii.edu}
\emailAdd{zhouyeling@ucas.ac.cn}

\abstract{
In the standard cosmological paradigm, the gravitational wave (GW) spectrum from a cosmic string network is well defined. However, early-Universe dynamics can significantly alter this relic signal. We systematically investigate how different new physics epochs in the radiation era—including extra relativistic degrees of freedom, early matter domination, intermediate vacuum domination, and kination—modify string network evolution and loop production. 
We find that a kination era enhances the stochastic GW signal, whereas the other scenarios suppress the signal to varying degrees. We also demonstrate that late-time dynamical dark energy has a negligible impact on the GW spectrum from cosmic strings.
}


\begin{document}

\thispagestyle{empty}
\def\thefootnote{\fnsymbol{footnote}}
\setcounter{footnote}{1}

\setcounter{page}{0}
\maketitle
\vspace{-1cm}
\flushbottom

\def\thefootnote{\arabic{footnote}}
\setcounter{footnote}{0}

%%%%%%%%%%%%%%%%%%%%%%%%%%%%%%%%%%
%			INTRODUCTION 					 %
%%%%%%%%%%%%%%%%%%%%%%%%%%%%%%%%%%
%\newpage

\section{Introduction} \label{sec:1}

Cosmic strings are one-dimensional objects of cosmic length and energy per unit length or tension $\mu$. They occur as topological defects during the spontaneous breaking of a continuous symmetry in an early Universe phase 
transition~\cite{Kibble:1976sj}. They arise naturally in grand unified theories that contain a gauged U(1) symmetry beyond the Standard Model (BSM)~\cite{Jeannerot:2003qv}. Cosmic strings also rise in superstring theory where microscopic fundamental strings are stretched to cosmological sizes by cosmic 
expansion~\cite{Sarangi:2002yt,Dvali:2003zj, Copeland:2003bj,Jackson:2004zg}. 

After they are produced, cosmic strings form a network in the expanding Universe. Long strings in the network intercommute with each other, producing sub-horizon scale loops chopped from long strings~\cite{Vilenkin:1984ib}. These loops oscillate and shrink gradually via gravitational radiation, providing a source for a stochastic gravitational wave (GW) background~\cite{Vilenkin:1981bx}. A key feature of the GW signal that makes this source distinguishable from other sources is a broadband spectrum of GWs from nHz to well above a Hz. In the framework of well-approximated Nambu-Goto strings in the Lambda Cold Dark Matter ($\LCDM$) model, the GW spectrum depends on only a single parameter $G\mu$, with $G$  the Newton constant; see Ref.~\cite{Schmitz:2024gds} for a recent review. 

As a time-integrated phenomenon, the stochastic GW background from cosmic strings chronicles the expansion history of the Universe. 
Consequently, any deviation from the standard $\LCDM$ paradigm alters the dynamics of the string network and the subsequent loop production which dictate GW emission. Ultimately, the early-Universe dynamics is imprinted onto the GW spectrum. 
While standard predictions typically adopt the spectrum from Nambu-Goto string simulations \cite{Blanco-Pillado:2011egf,Blanco-Pillado:2013qja,Blanco-Pillado:2017oxo}, which relies on the scaling solution in the radiation-dominated era \cite{Albrecht:1984xv,Bennett:1987vf,Allen:1990tv}, the insertion of a new physics (NP) epoch into the early Universe can distort this spectral shape by altering the scaling parameters characteristic of the NP era.

Given that cosmological observations have only constrained the cosmic history back to the epoch of Big Bang Nucleosynthesis (BBN), a modified cosmological evolution prior to BBN and post-inflation remains entirely viable. The primary objective of this work is to demonstrate that such uncertainties in the early-Universe history can induce substantial variations in the GW spectrum obtained from cosmic strings. To this end, we explore several well-motivated scenarios modifying the early radiation background, including models with extended BSM radiation degrees of freedom, an early epoch of matter or intermediate vacuum domination, and a kination-dominated phase.

%The suppressed intercommutation probability for superstrings allows a larger amplitude and accommodates the observed blue-tilted GW spectrum. For related work in the context of PTA data see Refs.~\cite{NANOGrav:2023hvm,Ellis:2023tsl, Ellis:2020ena, Blasi:2020mfx,Blanco-Pillado:2021ygr}. 
Because cosmic strings produce a GW spectrum over many orders of magnitude in frequency,
the numerous ongoing and planned GW observatories at different frequencies will be helpful to test the deviation of the GW spectrum from the conventional expected spectrum. 
Several groups of pulsar timing arrays (PTAs)~\cite{NANOGrav:2023gor,Antoniadis:2023ott,Reardon:2023gzh,Xu:2023wog} have reported hints of a stochastic GW background in the nHz band, although these are in tension with the GW spectrum from cosmic strings. 
GWs with frequencies above a mHz will be targeted by space-based laser interferometers (LISA~\cite{Audley:2017drz}, 
Taiji~\cite{Guo:2018npi}, TianQin~\cite{Luo:2015ght}, 
BBO~\cite{Corbin:2005ny}, DECIGO~\cite{Seto:2001qf}, 
$\mu$Ares~\cite{Sesana:2019vho}), atomic interferometers 
(MAGIS~\cite{Graham:2017pmn}, AEDGE~\cite{Bertoldi:2019tck}, 
AION~\cite{Badurina:2019hst}), and ground-based interferometers (Einstein Telescope~\cite{Sathyaprakash:2012jk} (ET), 
Cosmic Explorer~\cite{Evans:2016mbw} (CE), and Square Kilometer 
Array~\cite{Janssen:2014dka} (SKA).

The paper is organized as follows. In Section~\ref{sec:2}, we introduce simplified versions of NP epochs in the early Universe. We then investigate the corresponding string network evolution and loop production within the Nambu-Goto framework in Section~\ref{sec:3}. Section~\ref{sec:4} is devoted to the resulting GW signals across the different models and their comparisons with $\LCDM$. For completeness, in Section~\ref{sec:5} we evaluate the influence of dark energy models on the GW spectrum. Finally, we summarize our findings in Section~\ref{sec:6}.

\section{Modifications to cosmic evolution} \label{sec:2}

In the standard $\LCDM$ model, the flat Universe experiences eras of radiation domination~(RD), matter domination~(MD) and vacuum domination~(VD). 
The Hubble rate takes the form,
\begin{eqnarray} \label{eq:HLCDM}
H_{\rm \LCDM}(z) = H_0 \Big[\Omega_\Lambda + (1+z)^3 \Omega_{\rm m} + \mathcal{C}(z) (1+z)^4 \Omega_{\rm r}\Big]^{\frac12}\,,
\end{eqnarray}
where $\Omega_\Lambda = 1-\Omega_{\rm m} -\Omega_{\rm r}$ and 
\begin{eqnarray} \label{eq:C_eff}
\mathcal{C}(z) &=&
 \frac{g_* (z)}{g_*(0)} \left[\frac{h_*(z)}{h_*(0)}\right]^{-\frac43}
\end{eqnarray}
is a correction factor that accounts for the variation of the number of relativistic degrees of freedom (DOF) in the thermal plasma during the radiation era; $g_*$ and $h_*$ are the energy density and entropy degrees of freedom, respectively.
We apply a semi-analytical function introduced in Ref.~\cite{Marfatia:2023fvh} for the contribution of the Standard Model (SM) particles:
\begin{eqnarray} \label{eq:CSM_eff}
        \mathcal{C}_{\rm SM}(z)  = 1 - 0.05\sigma(29.30 + 8.73x) - 0.12\sigma(38.50 + 10.08x) 
        - 0.09\sigma(12.89 + 9.29x) \nonumber \\  
        - 0.28\sigma(10.14 + 12.70x) 
        - 0.042\sigma(-0.52 + 5.66x) - 0.03\sigma(-7.64 + 4.82x)\,,\quad
    \end{eqnarray}
where $x(z)=\ln\left[(1+z)\frac{T_0}{\rm GeV}\right]$
with $T_0$ the temperature of the Universe today,
and $\sigma(y) = 1/(1+e^{-y})$. 

If new physics contributes to the energy budget of the early Universe, the expansion history is modified.
This can be treated phenomenologically by inserting a new NP era during the RD era of the $\LCDM$ cosmology.
The Universe therefore evolves through the first radiation-dominated (RD1) era, the NP era, the second radiation-dominated (RD2) era, the MD era, and finally the VD era.
We assume that the NP era is short  compared to other eras.
Then, the Hubble expansion rate is simply parametrized as a  piecewise function of redshift,
\begin{align} \label{eq:HubbleNP}
    H(z)=\begin{cases}
        \displaystyle \frac{H_{\rm NP}(z_{\rm NP}^{\rm i})}{H_{\LCDM}(z_{\rm NP}^{\rm i})}\, H_{\LCDM}(z)  &  z_{\rm CS}^{\rm i} 
        \ge z>z_{\rm NP}^{\rm i} \\
        H_{\rm NP}(z) & z_{\rm NP}^{\rm i} \ge z > z_{\rm NP}^{\rm f}  \\
        H_{\LCDM}(z) & z_{\rm NP}^{\rm f} \ge z \ge 0 \,.
    \end{cases}
\end{align}
Here, $z_{\rm CS}^{\rm i}$ is the redshift in the RD era at which the string network forms. Thus, the history of the Universe for $z > z_{\rm CS}^{\rm i}$ is not relevant to us. The NP era lasts from an initial redshift $z_{\rm NP}^{\rm i}$ to a final redshift $z_{\rm NP}^{\rm f}$, and 
the Hubble parameter should be continuous at both these redshifts. 
The first line of Eq.~\eqref{eq:HubbleNP} ensures continuity of the Hubble parameter at $z=z_{\rm NP}^{\rm i}$, and the matching condition $H_{\rm NP}(z_{\rm NP}^{\rm f})=H_{\LCDM}(z_{\rm NP}^{\rm f})$ imposes continuity at the end of the NP era.
Before the NP era, i.e., $z>z_{\rm NP}^{\rm i}$, the Hubble expansion rate is that of the RD epoch in $\LCDM$, with its normalization  rescaled to match the NP era at $z=z_{\rm NP}^{\rm i}$.

In the following, we specify $H_{\rm NP}(z)$ for several early Universe scenarios. 
\begin{itemize}

\item {\bf Extra relativistic degrees of freedom (eDOF):} It is well known that extensions of the SM may introduce a large number of extra relativistic DOF at a certain new physics scale. For example, in  supersymmetric extensions of the SM, the number of DOF is at least double that of the SM. 
%Other BSM physics, such as the clockwork mechanism, Twin Higgs, and $N$naturaness model, and large DOFs in the dark sector, may introduce large DOFs above a certain BSM scale. 
In a Universe with $g_{*,{\rm SM}}+\Delta g_*$ and $h_{*,{\rm SM}}+\Delta h_*$ DOF, $\mathcal{C}_{\rm SM}(z)$ is replaced by
\begin{align}\label{dof}
    \begin{aligned}
        \mathcal{C}_{\rm eDOF}(z) = \mathcal{C}_{\rm SM}(z) \cdot\left(\frac{g_{*,{\rm SM}}(z)+ \Delta g_*}{g_{*,{\rm SM}}(z)}\right) \left(\frac{h_{*,{\rm SM}}(z)+ \Delta h_*}{h_{*,{\rm SM}}(z)}\right)^{-\frac43} \,,
    \end{aligned}
\end{align}
for $z>z_{\rm NP}^{\rm f}$. 
%In this model, the extra relativistic degrees of freedom are assumed to exist for $z>z_{\rm NP}^{\rm f}$.

\item {\bf Early matter domination (EMD):}
An early matter-dominated era may occur during, or before, the RD era.
This EMD era can be induced by heavy long-lived particles like cold dark matter~\cite{Moroi:1999zb}, axions~\cite{Nelson:2018via} and heavy neutrinos~\cite{Arcadi:2021doo}, as well as by primordial black holes~\cite{Green:2020jor}.
During the EMD era, the Hubble expansion rate is 
\begin{align} \label{eq:Hubble_EMD}
H_{\rm EMD}(z) = H_{\LCDM}(z_{\rm NP}^{\rm f}) \times \left(\frac{1+z}{1+z_{\rm NP}^{\rm f}}\right)^{3/2} \,,
\end{align}
where the normalization is fixed by matching to $H_{\LCDM}(z)$ at $z=z_{\rm NP}^{\rm f}$.
The duration of the EMD era, $z_{\rm NP}^{\rm f}<z<z_{\rm NP}^{\rm i}$, is determined by the mass and decay width of the long-lived matter component.

\item {\bf Intermediate vacuum domination (IVD):}
An intermediate vacuum-dominated era may arise during, or before, the RD era. Such a period may occur during thermal inflation~\cite{Lyth:1995ka} and supercooled phase 
transitions~\cite{Witten:1980ez}.
The Hubble parameter is approximately constant during the NP era:
\begin{align} \label{eq:Hubble_IVD}
    H_{\rm IVD}(z) = H_{\LCDM}(z_{\rm NP}^{\rm f}) \,.
\end{align}
The full expansion history is obtained by replacing $H_{\rm NP}(z)$ by $H_{\rm IVD}(z)$ in Eq.~\eqref{eq:HubbleNP}. 

\item {\bf Kination (KIN):}
During the kination era the kinetic energy of the fields dominates the energy density of the Universe, the equation-of-state parameter $w \equiv p/\rho=1$, and the energy density redshifts as $(1+z)^{-6}$; see Ref.~\cite{Gouttenoire:2021jhk} for a review.
If a kination era is embedded in the RD era, the Hubble parameter is
\begin{align} \label{eq:Hubble_KIN}
    H_{\rm KIN}(z) = H_{\LCDM}(z_{\rm NP}^{\rm f}) \times \left(\frac{1+z}{1+z_{\rm NP}^{\rm f}}\right)^3 \,.
\end{align}
%where the normalization is from matching to $H_{\LCDM}({\rm NP}^{\rm f})$. 
It is worth mentioning that although in our phenomenological approach we insert the kination period in the radiation era, it is more typical for kination to occur after a period of early matter domination, intermediate vacuum domination, or immediately after inflation.
We consider kination preceded by EMD or IVD at the end of Section~\ref{sec:4.2}.

\end{itemize}
The Hubble parameters in these early-Universe scenarios are summarized in Table~\ref{tab:Hubble}.

\begin{table}[t]  
	\centering 
    \begin{spacing}{1.5} 
	\begin{tabular}{c|c|c|c}  
		\hline\hline
        \multirow{2}{*}{}  & RD1 era & New Physics era & RD2+MD+VD eras \\
         & ($z_{\rm CS}^{\rm i} \ge z > z_{\rm NP}^{\rm i}$) & ($z_{\rm NP}^{\rm i}\ge z > z_{\rm NP}^{\rm f}$) & ($z\ge z_{\rm NP}^{\rm f}$) \\
        \hline
        $\LCDM$  &  \multicolumn{3}{c}{ $ \vphantom{\Big|} H_{\LCDM}(z)$ with $\mathcal{C}_{\rm SM}(z)$  } \\ 
        \hline
        eDOF  &  \multicolumn{2}{c|}{ $H_{\LCDM}(z)$ with $\mathcal{C}_{\rm eDOF}(z)$  } & $H_{\LCDM}(z)$ with $\mathcal{C}_{\rm SM}(z)$ \\ 
        \hline
        EMD & \multirow{3}{*}{ \raisebox{-1.5\height}{ \makecell{\!\!\!\!$ \frac{H_{\rm NP}(z_{\rm NP}^{\rm i})}{H_{\LCDM}(z_{\rm NP}^{\rm i})}\cdot H_{\LCDM}(z)$ \!\!\!\!\!\! \\ with $\mathcal{C}_{\rm SM}(z)$} } }   & $ \vphantom{\Bigg|} H_{\LCDM}(z_{\rm NP}^{\rm f}) \Big(\frac{1+z}{1+z_{\rm NP}^{\rm f}}\Big)^{3/2} $ &  \multirow{3}{*}{ \raisebox{-1.4\height}{ \makecell{$H_{\LCDM}(z)$ \\ with $\mathcal{C}_{\rm SM}(z)$} }  }  \\ 
        \cline{1-1} \cline{3-3}
        IVD &  &  $ \vphantom{\Bigg|}  H_{\LCDM}(z_{\rm NP}^{\rm f}) $ &  \\ 
        \cline{1-1} \cline{3-3}
        KIN &   &  $ \vphantom{\Bigg|} H_{\LCDM}(z_{\rm NP}^{\rm f}) \Big(\frac{1+z}{1+z_{\rm NP}^{\rm f}}\Big)^{3} $ &  \\ 
		\hline\hline
	\end{tabular}  	
    \end{spacing}	
	\caption{The Hubble parameter $H(z)$ in different scenarios. $H_{\LCDM}$ is given by Eq.~\eqref{eq:HLCDM} and $H_{\rm NP}$ is the Hubble parameter during the NP era; see Eqs.~\eqref{dof},~\eqref{eq:Hubble_EMD}, \eqref{eq:Hubble_IVD} and \eqref{eq:Hubble_KIN}. }
	\label{tab:Hubble}
\end{table}  

We model smooth transitions between epochs by replacing the step function by a sigmoid function ($\sigma(y)=1/(1+e^{-y})$):
\begin{align} \label{eq:sigma}
    \theta(z_2-z_1) \simeq \sigma\left[\frac{10}{\Delta N}\ln\left(\frac{1+z_2}{1+z_1}\right)\right] \,.
\end{align}
The sigmoid function increases from approximately 0 to 1 as its argument varies from $y=-5$ to $y=5$. 
The prefactor $10/\Delta N$ narrows the transition interval within $\Delta\ln(1+z)\sim \Delta N$, corresponding to a time interval $\Delta t\sim \Delta N \cdot H^{-1}$.
We fix $\Delta N=0.1$.
Thus, the sudden transition of the step function is smoothed over a short but finite interval, which stabilizes numerical calculations, and ensures that the transition between two eras completes within ${\cal O}(1)$ Hubble time.

\begin{figure}[t] 
	\begin{center} 
		\includegraphics[width=0.95\textwidth]{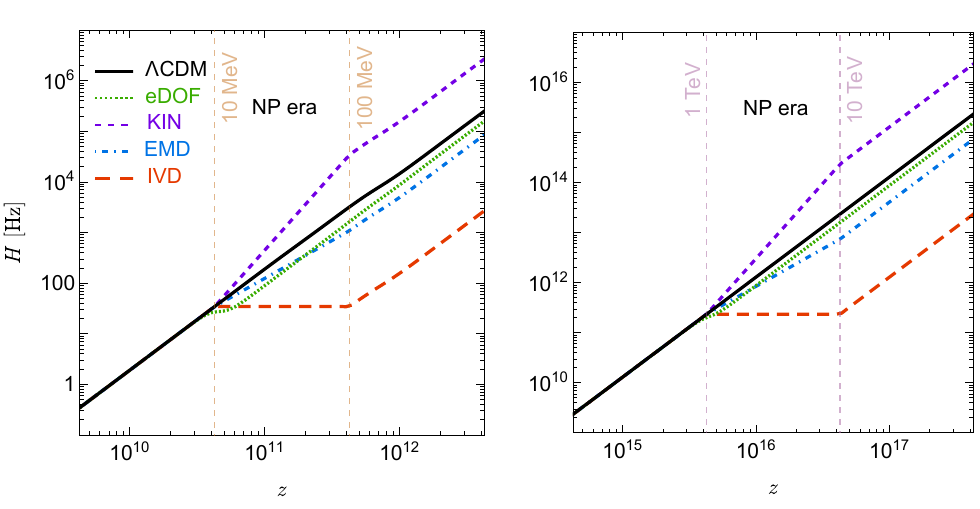}
		\caption{The Hubble parameter $H(z)$ for NP eras that span the energy scales [10~MeV, 100~MeV] (left panel) and [1~TeV, 10~TeV] (right panel). %The purple short-dashed, green dotted, blue dot-dashed, red long-dashed and black solid curves correspond to KIN, eDOF, EMD, IVD and $\LCDM$, respectively. 
        The vertical lines mark the corresponding redshift intervals of the  NP eras.}
		\label{fig:Hubble}
	\end{center}
\end{figure}

Except for the eDOF scenario, we assume that the NP era begins at an energy scale $10\Lambda_{\rm NP}$ and ends at $\Lambda_{\rm NP}$, which is equivalent to $1+z_{\rm NP}^{\rm i}=10(1+z_{\rm NP}^{\rm f})$.
For the eDOF scenario, the extra relativistic degrees of freedom are assumed to contribute throughout the epoch $z>z_{\rm NP}^{\rm f}$, and we take $\Delta g_*=\Delta h_*=10^3$.
The Hubble expansion rates for these scenarios are shown in Fig.~\ref{fig:Hubble}. For illustration, we consider two NP eras that span the energy scales, $[10~{\rm MeV},100~{\rm MeV}]$ and $[1~{\rm TeV},10~{\rm TeV}]$, corresponding to redshift ranges $[10z_{\rm NP}^{\rm f}, z_{\rm NP}^{\rm f}]$ with $z_{\rm NP}^{\rm f} \sim 4 \times 10^{10}$ (before BBN)  and $4 \times 10^{15}$, respectively.
These regimes are compatible with cosmological constraints from BBN, cosmic microwave background (CMB) measurements, and large-scale structure (LSS)~\cite{Hannestad:2004px,deSalas:2015glj}.
As can be seen from Fig.~\ref{fig:Hubble}, all the nonstandard cosmologies exhibit transition features near $\Lambda_{\rm NP}$ and $10\Lambda_{\rm NP}$  due to the sigmoid function in Eq.~\eqref{eq:sigma}.
The eDOF scenario has a different transition behavior because the extra relativistic degrees of freedom only affect $\mathcal{C}(z)$, so that $H(z)$ is proportional to the square root of the sigmoid function.
Thus, the transition in the eDOF scenario is smoother and extends over a slightly wider redshift interval in Fig.~\ref{fig:Hubble}.

\section{String network evolution} \label{sec:3}

We now describe the evolution of the cosmic string network and loop production in and beyond $\LCDM$.

\subsection{String dynamics and loop production} \label{sec:3.1}

%For Nambu-Goto strings, the energy evolution of free strings is given in \cite{Kibble:1984hp}
Neglecting energy-transfer mechanisms besides friction, the energy evolution of Nambu-Goto strings follows~\cite{Martins:1995tg} 
\begin{align}
    \begin{aligned} \label{eq:stringEOM_E}
        \frac{dE}{dt} &= H(1-2v^2)E - \frac{v^2}{\ell_f}E \,,
    \end{aligned}
\end{align}
where $\ell_f$ is the friction lengthscale and $v$ is the root-mean-square (rms) velocity of string segments.
The first term on the right-hand-side arises from the stretching and dilution of strings due to Hubble expansion, and the second term is due to friction from the retardation effect of the radiation fluid on the strings. 
The evolution of the energy density $\rho$ and the physical length $l$ of strings can be obtained by substituting $\rho\propto E/a^3$ and $l=E/\mu$ into Eq.~\eqref{eq:stringEOM_E}.
The evolution equation of the rms velocity is~\cite{Martins:1995tg}
\begin{align}
    \begin{aligned} \label{eq:stringEOM_v}
        \frac{dv}{dt} &= (1- v^2) \left[ \frac{k}{R} - 2 H v - \frac{v}{\ell_f} \right]\,,
    \end{aligned}
\end{align}
where $R$ is the effective 
curvature radius of the strings and $k$ is a phenomenological parameter related to their small-scale structure. 
The first term on the right-hand-side is the acceleration due to the curvature of the strings, and the second and third terms represent the deceleration via cosmic expansion and friction, respectively.
We neglect friction by assuming that $\ell_f$ is infinite.
Causality dictates that the rms velocity of the strings in Minkowski spacetime satisfy $v\le 1/\sqrt{2}$~\cite{Vilenkin:2000jqa}.
Note that Eqs.~\eqref{eq:stringEOM_E} and~\eqref{eq:stringEOM_v} apply to all string configurations including long strings and string loops, up to energy-transfer terms needed for specific string configurations.

The large-scale features of the string network are governed by the velocity-dependent one-scale (VOS) 
model~\cite{Martins:1995tg,Martins:1996jp}.
In the VOS model, the average curvature radius and average inter-string distance are equal and define a characteristic lengthscale $L$. On large scales, the energy density of the network is $\rho_\infty = \mu/L^2$ assuming one string segment of length $L$ in a volume $L^3$. Together with the average rms velocity of long strings $v_\infty$, the large-scale evolution of the string network is fully describable.
From Eq.~\eqref{eq:stringEOM_E}, the energy density of the string network satisfies
\begin{align}  \label{eq:drhodt}
    \frac{{d\rho}_\infty}{dt} &= - 2 H (1+v_\infty^2)\rho_\infty - \frac{ \tilde{c}\, v_\infty}{L} \rho_\infty  \,,
\end{align}
where the last term arises from the energy loss of the string network to string loops, and we have neglected the backreaction of loops on long strings. $v_\infty/L$ is the approximate collision rate of long strings and $\tilde{c}\simeq 0.23$ is the efficiency of chopping string loops from the network~\cite{Martins:2000cs}.
It is convenient to normalize the lengthscale $L$ as $\xi \equiv L/t$. Then, the evolution equations in Eqs.~\eqref{eq:stringEOM_v} and \eqref{eq:drhodt} simplify to 
\begin{align}
    \begin{aligned}
        t\,\dot{\xi} &= \beta (1+v_\infty^2) \xi - \xi + \frac{1}{2} \tilde{c}\, v_\infty \,, \\
        t\,\dot{v}_\infty &= (1- v_\infty^2) \left[ \frac{k(v_\infty)}{\xi} - 2 \beta v_\infty \right]\,,
    \end{aligned} \label{eq:VOS}
\end{align}
where $\beta = Ht = \{\frac12, \frac23, \frac13\}$ in the RD, MD and KIN eras, respectively, and $H$ is a constant in the VD era.
For $0\le v_\infty\le 1/\sqrt{2}$, the curvature parameter $k(v_\infty)$ can be expressed analytically as~\cite{Martins:2000cs} 
\begin{eqnarray} \label{eq:kv}
k(v_\infty) = \frac{2\sqrt{2}}{\pi} (1-v_\infty^2) (1+2\sqrt{2} v_\infty^3)\, \frac{1-8 v_\infty^6}{1+8 v_\infty^6} \,.
\end{eqnarray}
In this regime, long strings are stretched by cosmic expansion, and are effectively {\it infinitely} long.
For the values of $v_\infty$ of interest, $v_\infty \in [0.55,1/\sqrt2]$, $k(v) \simeq \frac{12}{\pi}(\frac{1}{\sqrt{2}}-v_\infty)$ is a good approximation. 

For constant $\beta$, Eq.~\eqref{eq:VOS} yields an attractor solution, called the scaling solution.
The scaling solutions in different eras, and the solution in the VD era, which does not follow a scaling law, are
\begin{align}
    \begin{aligned} \label{eq:scaling}
        \xi_{\rm r} &= 0.271\,, \;\quad v^{\rm r}_\infty = 0.662\,, \qquad \text{RD}  \\
         \xi_{\rm m} &= 0.625\,, \:\quad v^{\rm m}_\infty = 0.582\,, \qquad\, \text{MD} \\
        \xi_{\rm k} &= 0.155\,, \;\quad v^{\rm k}_\infty = 0.688\,, \qquad \text{KIN}  \\
        \xi_{\rm v} &\propto e^{Ht}/t\,, \;\quad v^{\rm v}_\infty \propto e^{-Ht}\,, \qquad ~\text{VD} \,.
    \end{aligned} 
\end{align}

\begin{figure}[t] 
	\begin{center} 
		\includegraphics[width=0.95\textwidth]{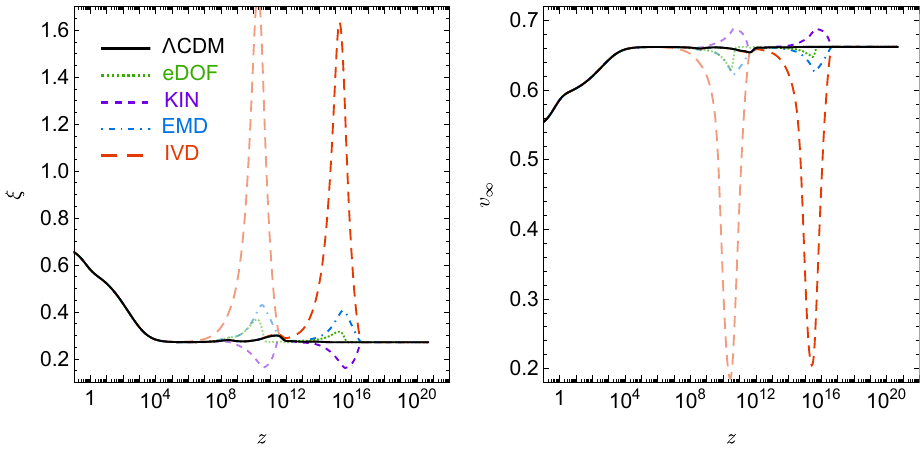}
		\caption{$\xi(z)$ and $v(z)$ in the VOS model with $G\mu=10^{-11}$ for the NP eras of Fig.~\ref{fig:Hubble} with the same colors and line types.
        The light and dark colors correspond to NP eras between [10~MeV, 100~MeV] and [1~TeV, 10~TeV], respectively.
        }
		\label{fig:xi_v}
	\end{center}
\end{figure}

In a dynamically evolving Universe, the scaling solutions in Eq.~\eqref{eq:scaling} may not hold. The evolution of $\xi(z)$ and $v(z)$ in the different scenarios is shown in Fig.~\ref{fig:xi_v}, which follows from the evolution of the Hubble parameter in Fig.~\ref{fig:Hubble}.
We take $G\mu=10^{-11}$, corresponding to the energy scale $\Lambda_{\rm CS}^{\rm i} \simeq 10^{8}~{\rm GeV} \times G\mu/{10^{-11}}=10^8$~GeV, and impose the initial conditions $\xi(t_{\rm CS}^{\rm i})=\xi_{\rm r}$ and $v_\infty(t_{\rm CS}^{\rm i})=v_\infty^{\rm r}$.
During the RD1 era, both $\xi$ and $v_\infty$ follow the scaling solutions of the RD era.
As the Universe enters the NP era, the network evolution deviates from the RD scaling solutions. 
According to the scaling solutions in Eq.~\eqref{eq:scaling}, $\xi$ increases and $v_\infty$ decreases when the Universe transits from RD1 to EMD or IVD.
In contrast, $\xi$ decreases and $v_\infty$ increases when the Universe transits from RD1 to KIN.
These tendencies are evident from Fig.~\ref{fig:xi_v}.
Since the NP era is short, the string network does not have time to evolve to the scaling regime in the NP era. 
After the NP era ends, the Universe returns to the RD2 era, and both $\xi$ and $v_\infty$ gradually approach the scaling solutions of the RD era.
Note that even in the RD era of $\LCDM$, small deviations from scaling appear for $10^8\lesssim z\lesssim 10^{12}$ because of the variation in the number of relativistic degrees of freedom encoded in $\mathcal{C}_{\rm SM}(z)$.
Similarly, in the eDOF scenario, $\xi$ and $v_\infty$ show only small deviations near $z_{\rm NP}^{\rm f}$, and then quickly return to the scaling regime.
As expected, whenever the Universe passes through a NP era, the string network temporarily deviates from the scaling solution.

We analyze the evolution of loops of length $l$ and number density $n(l,t)$. 
Loops are produced from long strings via collisions and self-bending. 
They lose energy via gravitational radiation with emission power $\frac{dE_\circ}{dt}\propto G\mu^2v_\circ^6$, where $E_\circ$ and $v_\circ$ are the energy and the root-mean-square velocity of loops. 
On including these energy losses and substituting $l=E_\circ/\mu$ in Eq.~\eqref{eq:stringEOM_E}, the loop length $l$ follows~\cite{Martins:1995tg} 
\begin{align} \label{eq:dldt_total}
    \frac{dl}{dt} = (1-2v_\circ^2)Hl - \Gamma G \mu \left(\sqrt{2}v_\circ\right)^6 \,,
\end{align}
where $\Gamma\simeq 50$ is the total GW emission power in units of $G\mu^2$ \cite{Blanco-Pillado:2017oxo} and the $\sqrt{2}$ normalization factor in the last term will become clear below.
Simulations find that string loops at production are much smaller than the Hubble horizon with $l/\pi \sim\mathcal{O}(1\%) \cdot H^{-1} $~\cite{Blanco-Pillado:2011egf, Blanco-Pillado:2013qja}.
Loops with velocity less than $1/\sqrt{2}$, are quickly stretched by cosmic expansion and become long strings according to Eq.~\eqref{eq:dldt_total}, while loops with velocity equal to $1/\sqrt{2}$ shrink by emitting GWs.
Therefore, loops smaller than the Hubble horizon satisfy $v_\circ^2=1/2$, and their length evolves according to 
\begin{align} \label{eq:dldt}
    \frac{dl}{dt}=-\Gamma G\mu \,.
\end{align}
The evolution of the energy density of loops $\rho_\circ$ ($\propto E_\circ/a^3$) is
\begin{align} \label{eq:drhocdt}
    \frac{{d\rho}_\circ}{dt} =& -3H\rho_\circ + \frac{{\cal F}}{\gamma_\circ}\cdot \frac{\tilde{c}\,v_\infty}{L} \rho_\infty - \Gamma G\mu^2 n(t) \,,
\end{align}
where $n(t) = \int n(l,t) \cdot dl$ is the total number density of loops at time $t$.
The second term on the right-hand-side arises from the assumption that all of the energy density lost by the network (via the last term in Eq.~\ref{eq:drhodt}) ends up in loops.
The Lorentz factor $\gamma_\circ = (1-v_{\circ}^2)^{-1/2}=\sqrt{2}$ accounts for energy loss from redshifting of the 
loop velocity~\cite{Martins:1996jp}.
The numerical factor ${\cal F} \sim \mathcal{O}(0.1)$ relaxes the assumption that all loops are created with exactly the same size, and is obtained by considering the distribution of 
loop lengths~\cite{Blanco-Pillado:2013qja}.
Given the number density of loops $n(l,t)\,dl$ with length in the interval $(l,l+dl)$ at time $t$, the energy density of loops is 
\begin{eqnarray} \label{eq:rhoc_flat}
    \rho_\circ = \int \mu l \cdot n(l,t)\, dl \,.
\end{eqnarray}

The instantaneous loop production function is defined as 
\begin{align} \label{eq:P_lt}
    \mathcal{P}(l,t) \equiv \frac{d n(l,t) }{dt}  \,,
\end{align}
so that ${\cal P}(l,t)\, dl \, dt$ is the increase in the number of loops with lengths in the interval $(l, l+dl)$ in the time period $(t, t+dt)$ per unit volume.
%It is different in different eras of Early Universe as will be discussed in the next subsection. 
To account for the Hubble expansion, we integrate ${\cal P}(l,t)$ from an initial time $t_{\rm CS}^{\rm i}$ to $t$ to obtain the loop number density distribution function,
\begin{eqnarray} \label{eq:n_lt}
    n(l,t) = \int_{t_{\rm CS}^{\rm i}}^t  \, {\cal P}(l'(t'),t') \left[\frac{a(t')}{a(t)}\right]^3 dt'\,,
\end{eqnarray}
where the loop length is obtained from Eq.~\eqref{eq:dldt}:
\begin{align} \label{eq:ltotal}
    l'(t') + \Gamma G \mu t'= l+ \Gamma G \mu t  \qquad t>t'\,,
\end{align}
and $t_{\rm CS}^{\rm i}$ is the time when friction becomes negligible~\cite{Gouttenoire:2019kij,Servant:2023tua,Schmitz:2024gds}.
Typically, the lower limit on the loop length is taken to be $l\gg \Gamma G\mu t$.

%%%%%%%%%%%%%%%%%%%%%%%%%%%%%%%%%%%%%
%%%%%%%%%%%%%%%%%%%%%%%%%%%%%%%%%%%%%
%%%%%%%%%%%%%%%%%%%%%%%%%%%%%%%%%%%%%

\subsection{Loop number density} \label{sec:3.2}

The loop number density $n(l,t)$, which depends on the loop production function $\mathcal{P}(l,t)$ in different eras of the early Universe, is essential for determining the GW spectrum observable today.  We begin with an evaluation of $\mathcal{P}(l,t)$ and $n(l,t)$ in $\LCDM$ and then generalize them for the NP eras.

In the RD era $\mathcal{P}(l,t)$ is well described by a monochromatic approximation and is compatible with simulations up to the efficiency factor ${\cal F}$ introduced in Eq.~\eqref{eq:drhocdt}~\cite{Blanco-Pillado:2013qja}.
Under this approximation, all loops are chopped with a length that is a constant fraction $\alpha_L$ of the correlation length $L$~\cite{Bennett:1985qt,Bennett:1986zn,Vilenkin:2000jqa}.\footnote{
In Refs.~\cite{Gouttenoire:2019kij,Gouttenoire:2021jhk}, it is assumed that loops are produced with lengths that scale with cosmic time as $l = \alpha t$. 
Our assumption that $l = \alpha_L L$ is more physical since $L$ is the characteristic lengthscale of the network. Then, $\alpha_L$ is related to $\alpha$ via $\alpha=\alpha_L\xi$;
see Refs~\cite{Auclair:2019wcv, Blanco-Pillado:2024aca} for a discussion of this issue. In the case of varying $\xi$, the choice of assumption makes a difference.}
The loop production function is given by
\begin{align} \label{eq:P_mono}
{\cal P}^{\rm mono}(l,t)= \frac{{\cal F}\tilde{c}}{\alpha_L\gamma_\circ} \cdot \frac{v_\infty(t)}{[\xi(t)]^4\, t^5} \cdot \delta \left( \alpha_L\xi(t) - \frac{l}{t}\right)\,,
\end{align}
where we have used the fact that 
\begin{eqnarray} \label{eq:drhocdt_flat}
    \int \mu l \cdot \frac{dn(l,t)} {dt} \, dl  = \frac{{\cal F}}{\gamma_\circ}\cdot \frac{\tilde{c}\,v_\infty}{L} \rho_\infty = \frac{{\cal F} \tilde{c}}{\gamma_\circ}\cdot \frac{v_\infty\mu}{\xi^3 t^3}\,.
\end{eqnarray}
Plugging Eq.~\eqref{eq:P_mono} into Eq.~\eqref{eq:n_lt} yields~\cite{Sousa:2013aaa, Auclair:2022ylu,Marfatia:2023fvh},
\begin{align} \label{eq:n_mono}
    \begin{aligned}
        n^{\rm mono}(l,t) &= \frac{{\cal F}\tilde{c}}{\alpha_L\gamma_\circ} \cdot \frac{v_\infty^\star}{\xi_\star^4\,t_\star^4} \, \frac{1}{|\alpha_L\xi_\star+\alpha_L\dot{\xi}_\star t_\star +\Gamma G\mu|}\,  \left[\frac{a_\star}{a(t)}\right]^3   \,.
    \end{aligned}
\end{align}
Here, $t_\star$ is the time when a loop is produced from the string network, so $\xi_\star \equiv \xi(t_\star)$, $v_\infty^\star \equiv v_\infty(t_\star)$ and $l_\star \equiv \alpha_L \xi_\star t_\star$ is fixed by the $\delta$ function.
One can determine $t_\star$ as a function of $l$ and $t$ by applying Eq.~\eqref{eq:ltotal} at $t'=t_\star$: 
\begin{equation}
(\alpha_L\xi_\star + \Gamma G\mu)t_\star = l + \Gamma G \mu t\,.
\end{equation}
Because Eq.~\eqref{eq:n_mono} holds for a nonzero $\dot{\xi}$, it allows for nonscaling effects.
In the scaling regime, $\dot{\xi}=0$, but if $Ht$ is time-dependent, $\xi_\star$ and $v_\infty^\star$ vary with time.
Thus, the loop number density in the scaling regime has $\dot{\xi}=0$ but time-varying $\xi_\star$ and $v_\infty^\star$:
\begin{align} \label{eq:n_mono_scaling}
        n^{\rm mono}_{\rm scaling}(l,t) &= \frac{{\cal F}\tilde{c}}{\alpha_L\gamma_\circ} \cdot \frac{v_\infty^\star}{\xi_\star^4\,t_\star^4} \, \frac{1}{(\alpha_L\xi_\star+\Gamma G\mu)}\,  \left[\frac{a_\star}{a(t)}\right]^3   \,.
\end{align}

In the radiation era, the loop number density with nonscaling effects, and specialized to the case of scaling, are respectively,
\begin{align}
    n^{\rm r}_{\rm nonscaling}(l,t) &\,= \frac{{\cal F}\tilde{c}}{\alpha_L\gamma_\circ} \cdot \frac{v_\infty^\star}{\xi_\star^4} \, \frac{1}{t^4} \frac{1}{|\alpha_L\xi_\star+\alpha_L\dot{\xi}_\star t_\star +\Gamma G\mu|}\,  \left[\frac{\alpha_L \xi_\star + \Gamma G\mu}{l/t + \Gamma G \mu}\right]^{\frac52}  \,,  \label{eq:n_r} \\
    n^{\rm r}_{\rm scaling}(l,t) &\,= \frac{{\cal F}\tilde{c}}{\alpha_L\gamma_\circ} \cdot \frac{v_\infty^\star}{\xi_\star^4}\frac{1}{t^4} \, \frac{(\alpha_L \xi_\star + \Gamma G\mu)^\frac32}{(l/t + \Gamma G \mu)^{\frac52}}  \,. \label{eq:n_r_scaling}
\end{align}
By fixing $\xi_\star$ and $v_\infty^\star$ to their scaling values $\xi_{\rm r}$ and $v_\infty^{\rm r}$ in Eq.~\eqref{eq:scaling}, we recover the result of the BOS simulation~\cite{Blanco-Pillado:2013qja},
\begin{align} \label{eq:n_r_BOS}
    n^{\rm r}_{\rm BOS}(l,t)&\, =\frac{0.18}{(l/t+\Gamma G\mu)^\frac52 t^4} \,, 
\end{align} 
for $\alpha_L \xi_{\rm r} \simeq 0.1 \gg \Gamma G\mu$.
Since $l/t$ is ${\cal O}(0.1)$ at production and monotonically decreasing with time,  $l/t < {\cal O}(0.1)$.

In the MD era, the loop production function suggested by the BOS simulation follows a broadband distribution rather than a monochromatic distribution, and in the scaling regime is given by~\cite{Blanco-Pillado:2013qja}
\begin{align} \label{eq:P_m_BOS}
{\cal P}_{\rm BOS}^{\rm m}(l,t)= \frac{0.141}{(l/t)^{1.69}\,t^5} \cdot \theta\left(0.18 - \frac{l}{t} \right)\,.
\end{align}
Thus, loops generated in the MD era have an upper limit on their length, $l/t\lesssim  0.18$.
The loop number density for a scaling network in the MD era is obtained from Eq.~\eqref{eq:n_lt} and the approximation $l'=l+\Gamma G\mu(t-t')\simeq l+\Gamma G\mu t$:
\begin{align} \label{eq:n_m_scaling}
n_{\rm scaling}^{\rm m}(l,t) \simeq \frac{0.27-0.45(l/t)^{0.31}}{(l/t + \Gamma G \mu)^2 \, t^4} \,,
\end{align}
with $l/t <0.18$~\cite{Blanco-Pillado:2013qja}. 

We attempt to incorporate the time dependence of $\xi$ and $v_\infty$ in the production function assuming that they vary slowly with time.
To do so, we make the assumption that the 
$(l/t)^{-1.69}$ and step function dependence continue to apply. Note  that this has not been validated by simulations.
Then, requiring that the loop production function in the matter era satisfies  Eq.~\eqref{eq:drhocdt_flat}, yields
\begin{align} 
        {\cal P}^{\rm m}(l,t) &= \frac{0.31}{(\alpha_L^{\rm m}\xi(t))^{0.31}} \frac{\mathcal{F}_{\rm m}\tilde{c}}{\gamma_\circ}\cdot \frac{1}{(l/t)^{1.69}} \frac{v_\infty(t)}{[\xi(t)]^3\, t^5} \cdot \theta\left(\alpha_L^{\rm m} \xi(t) - \frac{l}{t} \right)\,. \label{eq:P_m}
\end{align}
For Eq.~\eqref{eq:P_m} to match Eq.~\eqref{eq:P_m_BOS} with $\xi$ and $v_\infty$ fixed at their MD scaling values, we get $\mathcal{F}_{\rm m}\simeq 0.689$ and $\alpha_L^{\rm m}\xi_{\rm m}\simeq 0.18$.
The loop number density including nonscaling effects is found by inserting Eq.~\eqref{eq:P_m} in Eq.~\eqref{eq:n_lt}:
\begin{align} \label{eq:n_m}
    & n^{\rm m}_{\rm nonscaling}(l,t) = \frac{0.31}{(\alpha_L^{\rm m})^{0.31}} \frac{\mathcal{F}_{\rm m}\,\tilde{c}}{\gamma_\circ} \frac{1}{t^2} \int_{t_{\rm  eq}}^t \frac{v_\infty(t')}{[\xi(t')]^{3.31}} \frac{1}{l'^{1.69}} \frac{1}{t'^{1.31}}  \cdot \theta\left(\alpha_L^{\rm m} \xi(t') - \frac{l'}{t'} \right) dt'\,,
\end{align}
where $t_{\rm eq}$ is the epoch of matter-radiation equality.

%%%%%%%%%%%%%%%%%%%%%%%%%%%
%%%%%%%%%%%%%%%%%%%%%%%%%%%

In $\LCDM$, the total loop number density is
\begin{align}
    \begin{aligned} \label{eq:n_total_LCDM}
        n^{\rm total}_{\LCDM}(l,t) = \begin{cases}
            n^{\rm r}, &  z_{\rm CS}^{\rm i} \ge  z > z_{\rm eq}\,,  \\
            n^{\rm r\to m} + n^{\rm m}, & z_{\rm eq} \ge z > z_{\rm q}\,,  \\
            n^{\rm r\to v}  + n^{\rm m\to v} + n^{\rm v}, & z_{\rm q} \ge z \ge 0\,,
        \end{cases} 
    \end{aligned} 
\end{align}
where $z_{\rm eq}$ and $z_{\rm q}$ are the redshifts at matter-radiation equality and at which the expansion of the Universe started accelerating. They define the transitions from RD to MD and further to VD. Here, $n^{\rm m}$ ($n^{\rm v}$) is the number density of loops produced during the MD (VD) era. 
We denote the loops produced in the RD era that survive to the MD era by $n^{\rm r \to m}$~\cite{Ling:2025hjx}. Similarly, loops produced in the RD and MD eras that survive to the VD era are denoted by $n^{\rm r \to v}$ and $n^{\rm m \to v}$, respectively. 
In general, assuming that loops evolve independently of the network once produced, the number density of loops produced in a period ${\rm p}$ that survive to a later era $t$ is given by
\begin{align}
    \label{eq:n_p1_to}
        n^{\rm p\to}(l,t) &= n^{\rm p}\big(t_{\rm p}^{\rm f},l+\Gamma G\mu(t-t_{\rm p}^{\rm f})\big) \times \left(\frac{1+z}{1+z_{\rm p}^{\rm f}}\right)^3 \,, 
\end{align}
where $t_{\rm p}^{\rm f}$ and $z_{\rm p}^{\rm f}$ mark the end of the ${\rm p}$ period.
In the VD era, due to the exponential growth of the correlation length, loop production is negligible, ${\cal P}^{\rm v}(l,t)=0$, and
\begin{align} \label{eq:P_v}
    n^{\rm v}(l,t) = 0 \,.
\end{align}

%%%%%%%%%%%%%%%%%%%%%%%%%%%
%%%%%%%%%%%%%%%%%%%%%%%%%%%
%%%%%%%%%%%%%%%%%%%%%%%%%%%

Because the NP era is inserted in the RD era and after the formation of the string network, the network successively experiences the RD1, NP, RD2, MD, and VD eras.
New loops are produced from the string network and survive into later eras.
The total loop number density is a straightforward generalization of Eq.~\eqref{eq:n_total_LCDM}: 
\begin{align}
    \begin{aligned} \label{eq:n_total}
        n^{\rm total}(l,t) = \begin{cases}
            n^{\rm r1}, &  z_{\rm CS}^{\rm i} \ge z > z_{\rm NP}^{\rm i} \,, \\
            n^{\rm r1\to NP} + n^{\rm NP},  &  z_{\rm NP}^{\rm i} \ge z > z_{\rm NP}^{\rm f}\,, \\
            n^{\rm r1\to r2} + n^{\rm NP\to r2} + n^{\rm r2}, &  z_{\rm NP}^{\rm f} \ge z > z_{\rm eq}\,, \\
            n^{\rm r1\to m} + n^{\rm NP\to m} + n^{\rm r2\to m} + n^{\rm m}, & z_{\rm eq} \ge z > z_{\rm q}\,,  \\
            n^{\rm r1\to v} + n^{\rm NP\to v} + n^{\rm r2\to v} + n^{\rm m\to v}, & z_{\rm q} \ge z \ge 0\,.
        \end{cases} 
    \end{aligned} 
\end{align}
Here, $n^{\rm r1}$, $n^{\rm NP}$ and $n^{\rm r2}$ denote the number density of loops produced during the RD1, NP and RD2 eras, respectively.
For $z< z_{\rm NP}^{\rm f}$, the Universe evolves as in $\LCDM$, so $n^{\rm r2}$, $n^{\rm m}$, $n^{\rm r2 \to m}$, $n^{\rm r2 \to v}$ and $n^{\rm m \to v}$ are similar to $n^{\rm r}$, $n^{\rm m}$ $n^{\rm r\to m}$, $n^{\rm r\to v}$ and $n^{\rm m \to v}$ in Eq.~\eqref{eq:n_total_LCDM}, respectively. 
Loops produced in the RD1 and NP eras and that survive to later eras are 
included via $n^{\rm r1 \to NP}$, $n^{\rm r1 \to r2}$, $n^{\rm r1 \to m}$, $n^{\rm r1 \to v}$, $n^{\rm NP \to r2}$, $n^{\rm NP \to m}$, $n^{\rm NP \to v}$, and are 
calculated from $n^{\rm r1}$ and $n^{\rm NP}$ using Eq.~\eqref{eq:n_p1_to}. 
What remains is a derivation of $n^{\rm NP}$ in the different scenarios, to which we turn to shortly.

\begin{figure}[t]
	\centering
	\includegraphics[width=0.7\textwidth]{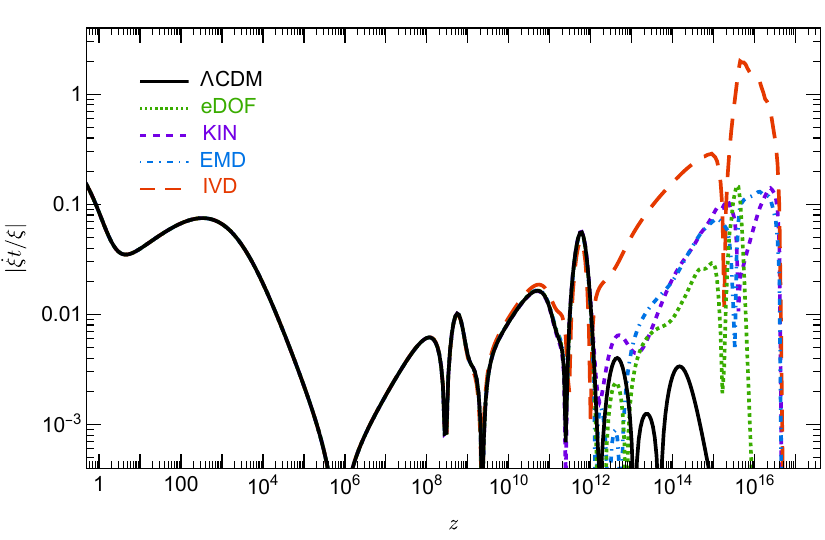}
    \caption{$|\dot{\xi} t/\xi|$ as a function of redshift for an NP era between [1~TeV, 10~TeV]. The colors and line types are the same as in Fig.~\ref{fig:Hubble}. }
    \label{fig:xidot}
\end{figure}

When the Universe transits from one era to another, the string network generally deviates from the scaling solution in the earlier era and subsequently approaches the scaling solution in the later era.
If the network deviates significantly from the scaling regime, nonscaling effects for loop production should be included.
We quantify the degree of deviation from scaling  by 
\begin{align} \label{eq:ratio_xi_dot} 
    \frac{\dot{\xi}t}{\xi} =  [Ht(1+v_\infty^2)-1] + \frac{\tilde{c}}{2}\frac{v_\infty}{\xi} \,,
\end{align}
where we have simply rearranged Eq.~\eqref{eq:VOS}. In Fig.~\ref{fig:xidot}, we plot this quantity as a function of redshift for $\LCDM$ and our NP scenarios. Clearly, $|\dot{\xi} t|$ is at most $\mathcal{O}(0.1\xi)$ except in the IVD era in which it is $\mathcal{O}(\xi)$. However, the degree of nonscaling in the IVD era is irrelevant because loop production during this era is vanishingly small. Thus, we neglect nonscaling effects on loop production.
As a check, in Appendix~\ref{app:A} we show that nonscaling effects on the GW spectrum are negligible. 
Henceforth, we assume the string network to always be in the scaling regime, even as it transits between different eras.

The efficiency of loop production depends on the background expansion. 
Although this dependence is not obvious from the collision terms in Eqs.~\eqref{eq:drhodt} and \eqref{eq:drhocdt}, it is reflected in the loop production functions in different cosmological eras, e.g., Eqs.~\eqref{eq:P_mono} and \eqref{eq:P_m_BOS}.
The greater the acceleration of background expansion, the more difficult it is to generate loops from the string network.
It is useful to characterize cosmic expansion through the deceleration parameter $q\equiv -\frac{\ddot{a}}{aH^2} = \frac1\beta -1$, where $q=-1$, $1/2$, $1$, and $2$ for the VD, MD, RD, and KIN eras, respectively.
Accelerated expansion with $q<0$ stretches the string network and suppresses loop formation, so that loop production is negligible; see Eq.~\eqref{eq:P_v}.
For the MD and EMD eras with $q = 1/2$, we use the broadband function to describe loop production, i.e., Eq.~\eqref{eq:P_m_BOS}.
For the RD and KIN eras with $q\ge 1$, we adopt the monochromatic approximation in Eq.~\eqref{eq:P_mono}.

We summarize the loop number density for our NP scenarios below. 
\begin{itemize}

    \item \textbf{eDOF.} The eras with $z>z_{\rm eq}$ are radiation dominated, so the loop number density is given by $n^{\rm total}_{\LCDM}$ in Eq.~\eqref{eq:n_total_LCDM}, except that ${\cal C}_{\rm eDOF}$ replaces ${\cal C}_{\rm SM}$ for $z>z_{\rm NP}^{\rm f}$.

        \item \textbf{KIN.} %As discussed in the last paragraph, we assume that the loop production in the KIN era follows the monochromatic approximation in Eq.~\eqref{eq:P_mono}. 
    Plugging $a(t) \propto t^{1/3}$ for the KIN era into Eq.~\eqref{eq:n_mono}, the number density of loops produced during the KIN era for a scaling network is
    \begin{align}
        n^{\rm NP}_{\rm scaling}(l,t) &\,= \frac{{\cal F}_{\rm k}\tilde{c}}{\alpha_L^{\rm k}\gamma_\circ} \cdot \frac{v_\infty^\star}{\xi_\star^4}\frac{1}{t^4} \, \frac{(\alpha_L^{\rm k} \xi_\star + \Gamma G\mu)^2}{(l/t + \Gamma G \mu)^3}  \,. \label{eq:n_KIN_scaling}
    \end{align}
    Here, $\mathcal{F}_{\rm k}$ and $\alpha_L^{\rm k}\xi_{\rm k}$ are factors to be determined by simulations. 
    Absent simulations in the kination era, solely for the purpose of illustration, we take the values in the RD era, $\mathcal{F}_{\rm k}\simeq 0.1$ and $\alpha_L^{\rm k}\xi_{\rm k}\simeq 0.1$. 

    \item \textbf{EMD.} Loop production in the EMD era follows $\mathcal{P}^{\rm m}_{\rm BOS}(l,t)$ in Eq.~\eqref{eq:P_m_BOS} for scaling networks. Hence, the number density of loops produced during the NP era is
    \begin{align}
        n^{\rm NP}_{\rm scaling}(l,t) = &\ n^{\rm m}_{\rm scaling}(l,t)  \simeq \frac{0.27-0.45(l/t)^{0.31}}{(l/t + \Gamma G \mu)^2 \, t^4}   \,.  \label{eq:n_EMD_scaling} 
    \end{align}

    \item \textbf{IVD.} The correlation length during the IVD era increases exponentially, similar to the VD era. Thus, $n^{\rm NP}=0$.

\end{itemize}

\begin{figure}[t]
	\centering
	\includegraphics[width=0.95\textwidth]{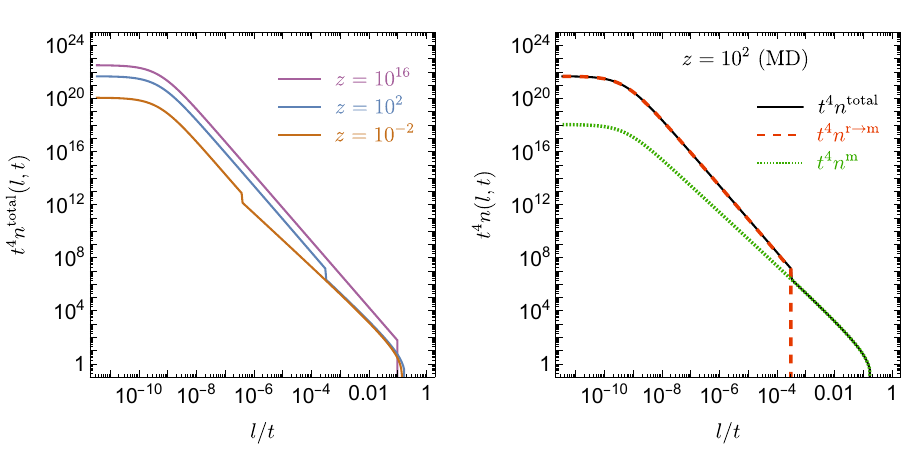}
    \caption{ Normalized total loop number density in $\LCDM$ in the RD, MD and VD eras (left panel), and its components in the MD era (right panel).
    In the right panel, the red dashed, green dotted and black solid curves correspond to loops produced in the RD era that survive to the MD era, loops produced in the MD era, and their total.
    We take $G\mu=10^{-11}$ so that $\Lambda_{\rm CS}^{\rm i}=10^8$~GeV, $\xi(t_{\rm CS}^{\rm i})=\xi_{\rm r}$ and $v_\infty(t_{\rm CS}^{\rm i})=v_\infty^{\rm r}$. }
    \label{fig:t4nloop_LCDM}
\end{figure}

In $\LCDM$, we take the loop number density in the RD and MD eras to be Eqs.~\eqref{eq:n_r_scaling} and \eqref{eq:n_m_scaling}, respectively, and neglect loop production in the VD era. The total loop number density
 as a function of the loop length is shown in the left panel of Fig.~\ref{fig:t4nloop_LCDM} for the same inputs as the black curves in Fig.~\ref{fig:xi_v}, and at three representative redshifts, $z=10^{16}, 10^2, 10^{-2}$, in the RD, MD and VD eras, respectively. Loop produced in the MD era and those that survive from the RD era are shown in the right panel for $z=10^2$.
The vertical steps in the loop production function correspond to the transition from the RD to MD era.
As noted in Ref.~\cite{Blanco-Pillado:2013qja}, loops of a given length $l$ at time $t$ in the MD and VD eras are dominated by loops produced in the RD era.

\begin{figure}[t]
	\begin{center} 
		\includegraphics[width=0.95\textwidth]{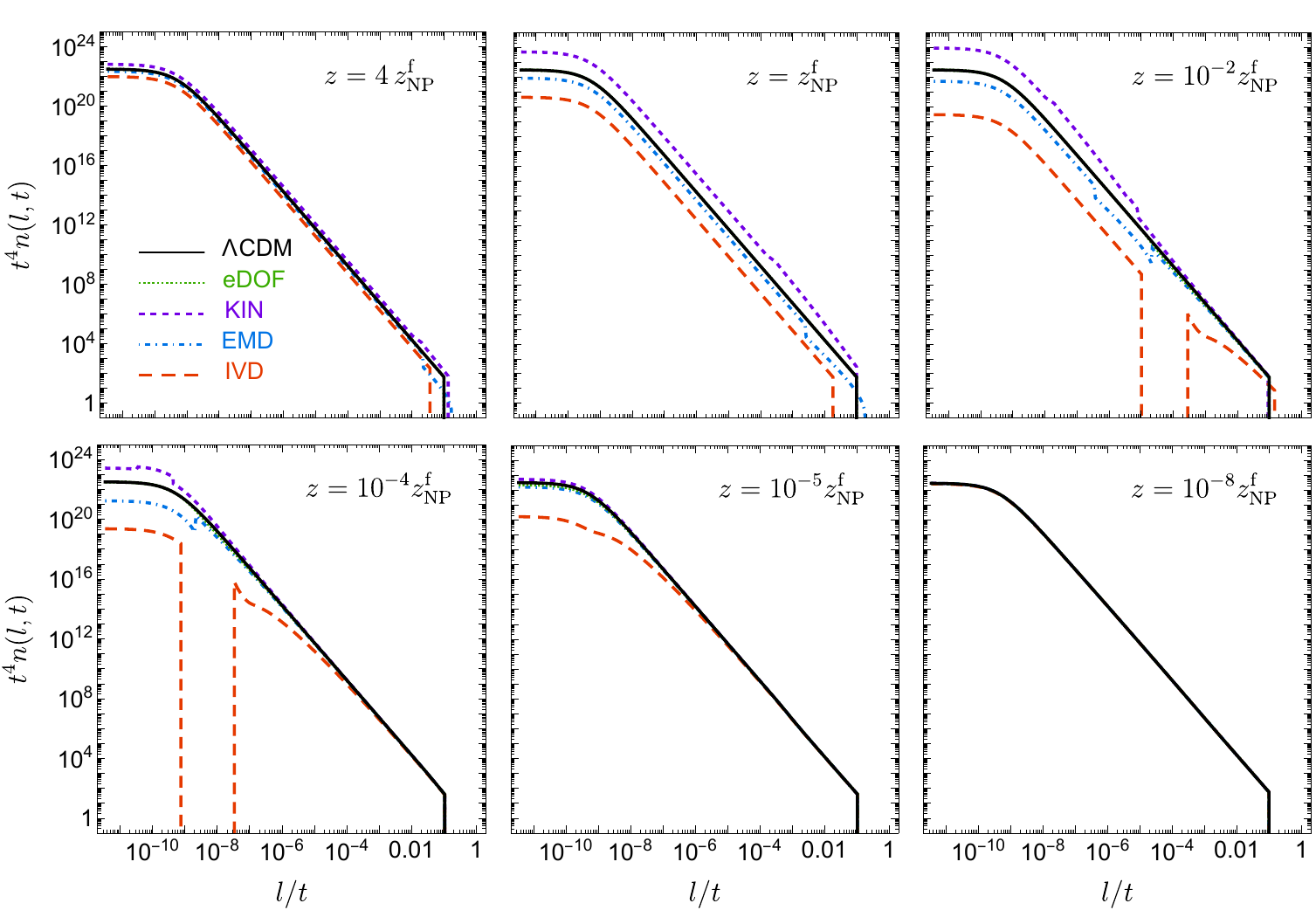}
		\caption{ Normalized loop number density at different redshifts for an NP era between [1~TeV, 10~TeV], corresponding to $z_{\rm NP}^{\rm f}=4.26\times 10^{15}$. The colors and line types are the same as in Fig.~\ref{fig:Hubble}. }
		\label{fig:t4nloop}
	\end{center}
\end{figure}

We show the evolution of $t^4 n(l,t)$ with redshift in Fig.~\ref{fig:t4nloop} for an NP era between [1~TeV,~10~TeV].
For the loop number density in Eq.~\eqref{eq:n_total}, $n^{\rm r1}$ and $n^{\rm r2}$ are given by Eq.~\eqref{eq:n_r_scaling}, $n^{\rm m}$ by Eq.~\eqref{eq:n_m_scaling}, and $n^{\rm NP}$ takes the values 0, Eq.~\eqref{eq:n_EMD_scaling}, and Eq.~\eqref{eq:n_KIN_scaling} in the IVD, EMD, and KIN eras, respectively.
Before the NP era, the loop number density is $n^{\rm r1}$ in the different scenarios.
After the Universe enters the NP era, the majority of loops are those produced in the RD1 era for most $l/t$, and new loops produced in the NP era have large $l/t$.
The evolution of $t^4n^{\rm r1\to NP}$ and $t^4n^{\rm NP}$ in  the different scenarios is different except for the eDOF scenario. 
The differences in $t^4n(l,t)$ in the NP eras get larger towards the end of the NP era ($z=z_{\rm NP}^{\rm f}$) as is evident from Fig.~\ref{fig:t4nloop}.
In the initial phase of the RD2 era, $t^4n^{\rm r2}$, $t^4n^{\rm r1\to r2}$ and $t^4n^{\rm NP\to r2}$ differ across scenarios.
After $\xi_\star$ and $v_\infty^\star$ in the RD2 era reach their scaling values in Eq.~\eqref{eq:scaling}, $t^4n^{\rm r2}$ is the same in all scenarios.
After the network evolves for a few orders of magnitude in redshift, the new loops produced in the RD2 era dominate the number density, and $t^4n(l,t)$ becomes the same in all the scenarios; see the bottom-right panel of Fig.~\ref{fig:t4nloop}.
Note that the dip (bump) in the bottom-left panel in the EMD (KIN) scenario is because the number density of loops produced during the EMD (KIN) era is much smaller (larger) than that during the RD2 era. 
The huge break in $t^4n(l,t)$ for IVD is explained by the fact that no loops are produced during the IVD era. The break moves to the left with time (decreasing redshift), and the left and right hand edges of the break correspond to loops produced in the RD1 and RD2 eras, respectively. The suppression in $t^4n(l,t)$ at the beginning of the RD2 era is because $\xi_\star$ is larger and $v_\infty^\star$ is smaller than their scaling values in Eq.~\eqref{eq:scaling}. 
Compared to $\LCDM$, the KIN scenario enhances $t^4n$, while the eDOF, EMD and IVD scenarios suppress $t^4n$ in increasing order.

To further analyze the evolution of $t^4n(l,t)$ with redshift, we consider loops produced in a period p that survive to a later era.
%and define $t^4n^{\rm p}(l,t)=t^4n^{\rm p}(x)$ and $t^4n^{\rm p\to}(l,t)=t^4n^{\rm p\to}(x)$ with $x=l/t$ after neglecting the varying of $\xi$ and $v_{\infty}$.
From Eqs.~\eqref{eq:n_r_scaling}, \eqref{eq:n_m_scaling}, \eqref{eq:n_EMD_scaling} and \eqref{eq:n_KIN_scaling}, we find $t^4n^{\rm p}(l,t) \propto (l/t+\Gamma G\mu)^{3\beta-4}$,
where $\beta=\frac12,\frac23,\frac13$ in the RD, MD and KIN eras, respectively.
For loops that survive to later eras, $n^{\rm p\to}(l,t)\propto {(t_{\rm p}^{\rm f})^{-3\beta}}\frac{(a_{\rm p}^{\rm f})^3}{a^3} (l+\Gamma G\mu t)^{3\beta-4}$ according to Eq.~\eqref{eq:n_p1_to}. Thus, at time $t$,
\begin{align}
    \frac{t^4n^{\rm p\to}(l,t)}{t_{\rm p}^4n^{\rm p}(l,t)} = \left(\frac{t}{t_{\rm p}^{\rm f}}\right)^{3\beta} \left( \frac{1+z}{1+z_{\rm p}^{\rm f}} \right)^3\,, \label{eq:ratio_t4nloop}
\end{align}
where $t_{\rm p}$ is any time in period p.
We find that Eq.~\eqref{eq:ratio_t4nloop} matches the numerical results in Figs.~\ref{fig:t4nloop_LCDM} and~\ref{fig:t4nloop} to within $\mathcal{O}(1\%)$.

%%%%%%%%%%%%%%%%%%%%%%%%%%%%%%
%%%%%%%%%%%%%%%%%%%%%%%%%%%%%%
%%%%%%%%%%%%%%%%%%%%%%%%%%%%%%

\section{Gravitational waves}
\label{sec:4}

Loops chopped from the string network are the main source of GWs.
 We briefly review how GWs are produced from string loops in  $\LCDM$ in Section~\ref{sec:4.1}, and extend the formalism to the NP scenarios in Section~\ref{sec:4.2}.

\subsection{GWs from string loops} \label{sec:4.1}
The spectrum of the gravitational-wave background as a function of frequency $f$ in the present epoch $t_0$ is given by
\begin{eqnarray}
h^2 \Omega_{\rm GW} (f)  = \frac{h^2}{\rho_c} \frac{d\rho_{\rm GW}(t_0, f)}{d\ln f}\,, \label{eq:h2Omega_define}
\end{eqnarray}
where $d\rho_{\rm GW}(t_0,f)/{d\ln f}$ is the energy density of GWs per logarithmic frequency interval and $\rho_c = 3 H_0^2 /(8\pi G)$ is the critical energy density. 
 Because the relic GW background today is sourced from gravitational radiation of frequency $f'$ emitted at time $t$, the redshift of the frequency and the $(1+z)^{-4}$ dilution of the energy density due to the expansion of the Universe must be taken into account.
The energy density of the relic GW background today can be written as
\begin{eqnarray} \label{eq:rho_GW}
\frac{d \rho_{\rm GW}(t_0, f)}{df}
= \int_{t_{\rm CS}^{\rm i}}^{t_0} \frac{dt}{[1+z(t)]^4} P_{\rm GW} (t,f') \frac{\partial f'}{\partial f}\,, 
\end{eqnarray}
where $f'=2k/l$ corresponds to the emission frequency from a loop of length $l$ in harmonic mode $k =1,2,3\ldots$, and $P_{\rm GW}(t,f')$ is the GW power emitted at time $t$ at frequency $f'=(1+z)f$,
\begin{eqnarray} \label{eq:P_GW}
    P_{\rm GW} (t,f') = G\mu^2 \sum_{k=1}^{\infty} \frac{l}{f'} n(l,t) P_k \,.
\end{eqnarray} 
%It plays a crucial role in the determination of the GW spectrum, and we will revisit its implications throughout the remainder of the paper. 
Here, $P_k$ is the normalized GW power (in units of $G\mu^2$) emitted by a loop in harmonic mode $k$.  For GW emission via cusps, kinks, and kink-kink collisions, 
\begin{eqnarray}\label{eq:P_k}
P_k \simeq \frac{\Gamma k^{-q}}{\zeta(q)} \,,
\end{eqnarray}
with $q=4/3$, 5/3 and 2 for cusps, kinks and kink-kink collisions, and $\zeta(q) = \sum_{k=1}^\infty k^{-q}$ is the Riemann zeta function. 
We only include the contribution from cusps as they are the dominant source of GWs.

%%%%%%%%%%%%%%%%%%%%%

\begin{figure}[t]
	\centering
	\includegraphics[width=0.8\textwidth]{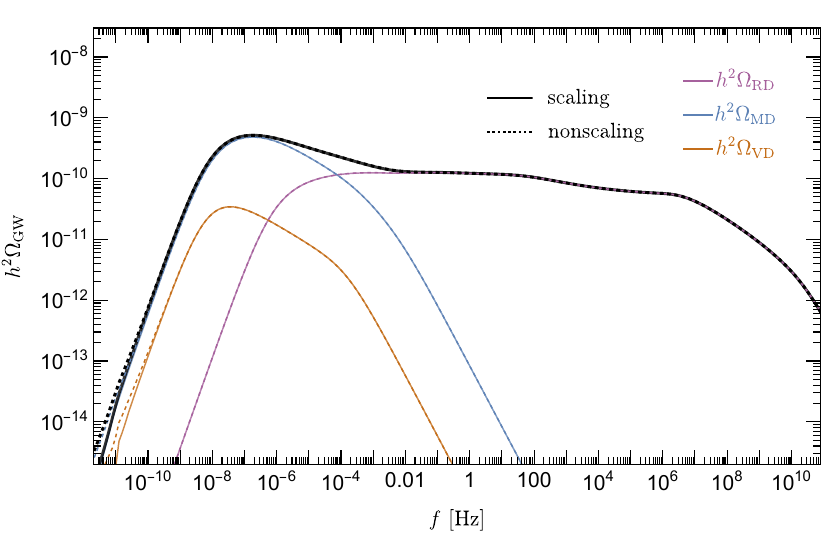}
	\caption{ GW spectra in $\LCDM$ for the same inputs as in Fig.~\ref{fig:t4nloop_LCDM}. The solid and dotted curves are the scaling and nonscaling solutions. The purple, blue, brown and black curves correspond to GWs produced during the RD, MD and VD eras and their total, respectively.  
    }
	\label{fig:GW_spectrum_LCDM}
\end{figure}

Putting all this together, and replacing time by redshift as the integration variable $\big(\int_{t_{\rm CS}^{\rm i}}^{t_0}dt = \int_0^{z_{\rm CS}^{\rm i}} H(z)^{-1} dz/(1+z)\big)$, we arrive at the integrated GW spectrum:
\begin{eqnarray} \label{eq:h2Omega_k}
h^2\Omega_{\rm GW} (f) = \sum_{k=1}^{\infty} h^2 \Omega^{(k)}_{\rm GW} (f)\,,
\end{eqnarray}
where the contribution of each harmonic mode is
\begin{eqnarray} \label{eq:h2Omegak}
 h^2 \Omega^{(k)}_{\rm GW} (f) = \frac{8\pi}{3} \left( \frac{h}{H_0} \right)^2
(G\mu)^2 \, P_k \, \frac{2k}{f} \int_0^{z_{\rm CS}^{\rm i}} \frac{dz}{(1+z)^6 H(z)} \, n\left(\frac{2k}{(1+z)f}, t(z)\right)\,.
\end{eqnarray}
 The harmonic modes should in principle be summed up to infinity, but summing up to $k=10^4$ keeps the relative error below $4 \%$, which is adequate~\cite{Marfatia:2023fvh}. 

The GW spectrum in $\LCDM$ for $\Lambda_{\rm CS}^{\rm i}=10^8$~GeV is shown in Fig.~\ref{fig:GW_spectrum_LCDM}. 
GWs produced in the RD, MD and VD eras are shown separately. 
The solid curves are the scaling solutions in Eqs.~\eqref{eq:n_r_scaling} and~\eqref{eq:n_m_scaling}, and the dotted curves are the nonscaling solutions in Eqs.~\eqref{eq:n_r} and \eqref{eq:n_m}. They are almost indistinguishable.

\subsection{Impact of cosmic evolution on the GW spectrum} \label{sec:4.2}

%\subsection{GWs from loops in different models of Early Universe}
We apply Eqs.~\eqref{eq:h2Omega_k} and~\eqref{eq:h2Omegak} to calculate the GW spectra in the four NP scenarios. 
The Hubble parameter in each scenario is provided in Table~\ref{tab:Hubble}, and the loop number density in the different eras is detailed  in Section~\ref{sec:3.2}.

\begin{figure}[t]
	\centering
		\includegraphics[width=0.95\textwidth]{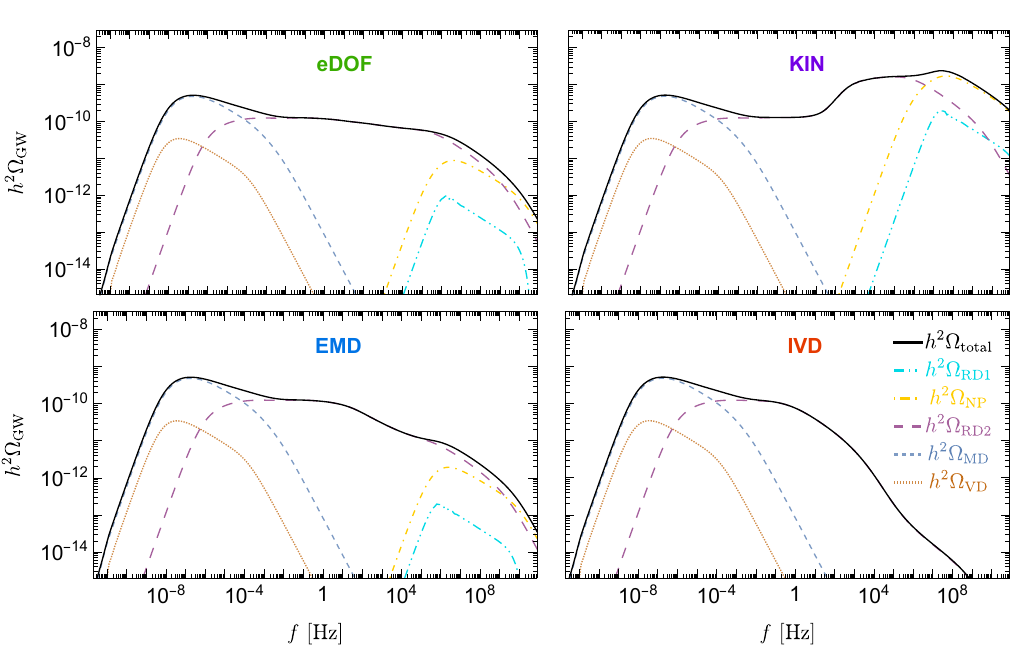}
		\caption{ GW spectra for each NP era between [1~TeV, 10~TeV] and $G\mu=10^{-11}$.
        For each scenario, the GW spectrum produced in different epochs is shown separately.     }
		\label{fig:GW_spectrum_contribution}
\end{figure}

\begin{figure}[t]
	\centering
		\includegraphics[width=0.8\textwidth]{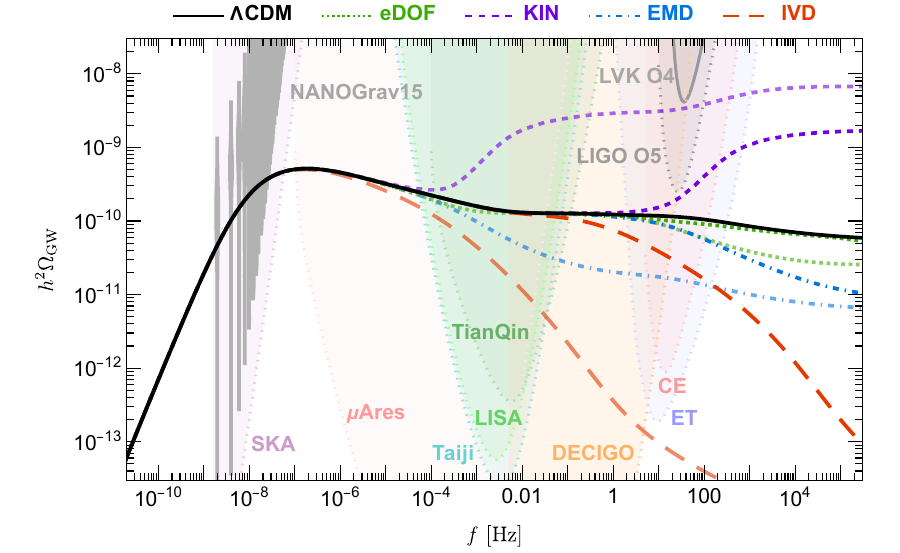}
		\caption{ 
        A comparison of the GW spectra for NP eras between [10~MeV, 100~MeV] (light colors) and [1~TeV, 10~TeV] (dark colors), and $G\mu=10^{-11}$.
        The purple short-dashed, green dotted, blue dot-dashed, red long-dashed and black solid curves correspond to KIN, eDOF, EMD, IVD and $\LCDM$, respectively. Experimental sensitivity curves are also shown.
        }
        \label{fig:GW_spectrum_experiments}
\end{figure}

In Fig.~\ref{fig:GW_spectrum_contribution} we show GW spectra for each NP era between [1~TeV, 10~TeV]. For each scenario, the GW spectrum produced in different epochs is shown separately.   
A comparison of the spectra for NP eras between [10~MeV, 100~MeV] and [1~TeV, 10~TeV] is displayed in Fig.~\ref{fig:GW_spectrum_experiments}.
Correspondingly, deviations from the spectrum for $\LCDM$ appear at  frequencies above $10^{-4}$~Hz, and 10~Hz, respectively. 
At high frequencies, $\Omega_{\rm GW}^{\rm KIN}>\Omega_{\rm GW}^{\LCDM}\gtrsim\Omega_{\rm GW}^{\rm eDOF}>\Omega_{\rm GW}^{\rm EMD}>\Omega_{\rm GW}^{\rm IVD}$, in proportion to the loop number densities produced in the corresponding periods. In the $[0.1~{\rm mHz}, 0.1~{\rm Hz}]$ band favored by the space-based laser interferometers LISA, Taiji and TianQin, all four scenarios have similar spectra to $\LCDM$ if the NP scale is above a TeV. However, for TeV-scale NP, the spectra $\gtrsim {\rm Hz}$ deviate sufficiently from
$\LCDM$ (except for eDOF) and from each other to be easily differentiated by the ground-based laser interferometers ET and CE.
Quantitative measures to discriminate between the GW spectra are the signal-to-noise ratio and Prominence~\cite{Goncalves:2025lat}. 
Note that for NP eras before BBN, the GW spectra in the PTA band are independent of the NP scenario.

\begin{figure}[t]
	\centering
		\includegraphics[width=0.8\textwidth]{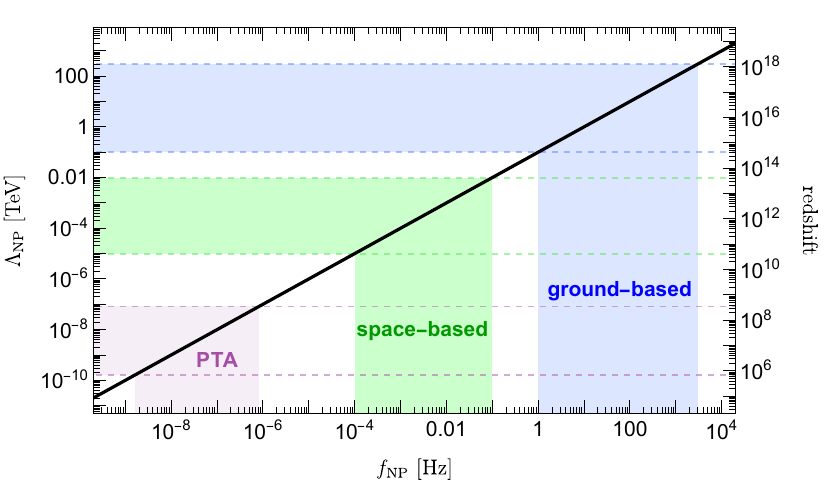}
		\caption{The NP energy scale probed by GWs of a characteristic frequency produced by cosmic strings. The purple, green and blue regions indicate the frequency ranges accessible to PTAs, future space-based interferometers, and future ground-based interferometers, respectively.
        }
        \label{fig:frequency_scale}
\end{figure}

For a given NP scale $\Lambda_{\rm NP}$, we estimate the lowest frequency $f_{\rm NP}$ at which the NP GW spectrum deviates from the spectrum for $\LCDM$. 
Note that the following analysis does not apply to the IVD scenario because its loop number density in Fig.~\ref{fig:t4nloop} changes drastically with redshift.
From Eq.~\eqref{eq:h2Omegak}, GWs emitted during the RD2 era satisfy
\begin{align*}
    \frac{d\Omega^{(k)}_{\rm GW} (f)}{dz} \propto  \frac{n(l, t)}{(1+z)^8} \propto t^4n(l, t)  \,,
\end{align*}
The maximum of $t^4n(l,t)$ occurs for $l/t\sim \Gamma G\mu$, as can be seen from Eq.~\eqref{eq:n_r_scaling}.
Thus, for loops produced at time $t_\star$ with initial length $l_\star$, the redshift at which $t^4n(l,t)$ is maximum is $z(t) \sim z(t_\star) \sqrt{\frac{2\Gamma G\mu}{l_\star/t_\star}}$~\cite{Gouttenoire:2021jhk}.
For redshifts below $10^{-4}z_{\rm NP}^{\rm f}$ and $l/t\sim 10^{-10}$, the difference in $t^4n(l,t)$ between $\LCDM$ and the NP scenarios falls rapidly. 
Thus, GWs of frequency $f_{\rm NP}$ are emitted at $10^4z_{\rm NP}^{\rm f}$ by loops produced at $t_{\rm NP}^{\rm f}$, and $f_{\rm NP}$ is given by
\begin{align} \label{eq:f_NP}
    f_{\rm NP} \sim \frac{1}{1+10^{-4}z_{\rm NP}^{\rm f}} \frac{1}{l_\star(z_{\rm NP}^{\rm f})}  \sim  (10~{\rm Hz}) \times \frac{\Lambda_{\rm NP}^{\rm f}}{{\rm TeV}}  \,,
\end{align}
where we have used $l\sim l_\star(z_{\rm NP}^{\rm f}) \sim \frac{\alpha_L\xi_{\rm r}}{H(z_{\rm NP}^{\rm r})}$, and $\Lambda_{\rm NP}^{\rm f}$ is the energy scale at the end of the NP era.
If $\Lambda_{\rm NP}^{\rm f} = 10$~MeV (1~TeV), the GW spectrum starts to deviate from the standard one at about $f_{\rm NP} \sim 10^{-4}$~Hz (10~Hz), consistent with Fig.~\ref{fig:GW_spectrum_experiments}. 
The relation between the characteristic frequency $f_{\rm NP}$ and the energy scale of the NP era is shown in Fig.~\ref{fig:frequency_scale}.
The purple, green and blue regions indicate the frequency ranges 
accessible to PTAs, future space-based interferometers, and future ground-based interferometers, respectively.
%The cosmological constraints from BBN, CMB and LSS \cite{Hannestad:2004px,deSalas:2015glj} require $\Lambda_{\rm NP}\gtrsim 5$~MeV.

\begin{figure}[t]
	\centering
    \begin{minipage}[t] {\textwidth}
	\begin{center} 
		\includegraphics[width=0.95\textwidth]{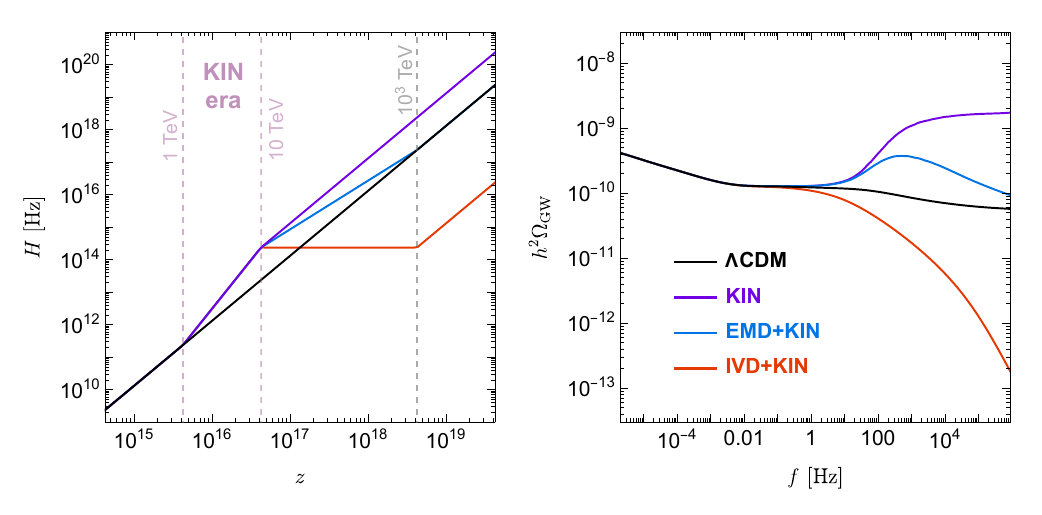}
		\caption{Hubble parameters and GW spectra for the EMD+KIN and IVD+KIN scenarios compared with KIN and $\LCDM$. We take $G\mu=10^{-11}$. The KIN era is between [1~TeV, 10~TeV] (light purple dashed lines), while the EMD and IVD eras span [10~TeV, $10^3$~TeV]. }
		\label{fig:H_xi_v_KIN}
	\end{center}
    \end{minipage} 
\end{figure}

In realistic models that have a KIN era, an era of EMD or IVD triggers 
kination~\cite{Spokoiny:1993kt,Joyce:1996cp, Co:2019wyp, Gouttenoire:2021jhk}. We denote such scenarios as EMD+KIN and IVD+KIN, respectively. In the EMD+KIN scenario, if a significant amount of radiation is not injected during either era, i.e., entropy injection is absent, and if the effective number of relativistic DOF is unchanged, then $\Lambda_{\rm EMD}^{\rm i}/\Lambda_{\rm EMD}^{\rm f}=(\Lambda_{\rm KIN}^{\rm i}/\Lambda_{\rm KIN}^{\rm f})^2$, with $\Lambda_{\rm EMD}^{\rm f}=\Lambda_{\rm KIN}^{\rm i}$,  where $\mathrm{i}$ and $\mathrm{f}$ denote the beginning and end of each era, respectively~\cite{Gouttenoire:2021jhk}. 
We impose this condition on the IVD+KIN scenario as well.
In Fig.~\ref{fig:H_xi_v_KIN}, we compare the Hubble parameters and GW spectra of the EMD+KIN and IVD+KIN scenarios with those of KIN and $\LCDM$. Here, the KIN era spans the energy range [1~TeV, 10~TeV] while the EMD and IVD eras span 
[10~TeV, $10^3$~TeV].
The transition from these eras to kination modifies the GW spectrum in the high frequency band, and makes it different from the spectrum in the KIN scenario considered so far.

\section{GWs in dark energy models} \label{sec:5}
For completeness, we quantify the extent to which the GW spectrum from cosmic strings is affected by relaxing the assumption that the accelerated expansion of the Universe is described by a cosmological constant $\Lambda$ with an equation-of-state parameter $w\equiv p/\rho =-1$. We allow for dynamical dark energy (DDE) with a redshift-dependent $w$, as hinted by recent data~\cite{DESI:2024mwx,DESI:2025zgx,DESI:2025fii,DES:2025upx}. 

The Hubble parameter in a spatially flat Universe is written as
\begin{align} \label{eq:HDDE}
    H_{\rm DDE}(z) = H_0 \left[  f(z)\Omega_{\rm DE} + (1+z)^3\Omega_{\rm m} + \mathcal{C}_{\rm SM}(z)(1+z)^4\Omega_{\rm r} \right]^{\frac12} \,,
\end{align}
where $f(z)\equiv \frac{\rho_{\rm DE}(z)}{\rho_{\rm DE}(0)}$ describes the redshift evolution of the dark energy density.
For a general dynamical dark energy component, the Friedmann equation gives
\begin{align} \label{eq:fDE}
    f(z) = \exp\left[3\int_0^z[1+w(z')]\frac{dz'}{1+z'}\right] \,.
\end{align}
The cosmological constant corresponds to the special case $w(z)=-1$, and hence $f(z)=1$.

\begin{table}[t]  
	\centering 
    \begin{spacing}{1.3} 
	\begin{tabular}{c|c|c|c}  
		\hline\hline
        Model  & $w(z)$ & $f(z)$ & parameter values \\
        \hline 
        $\Lambda$ & $-1$ & $1$ &  \\
        \hline
        CPL  & $w_0 + w_a\frac{z}{1+z}$ & $(1+z)^{3(1+w_0+w_a)}e^{-3w_a\frac{z}{1+z}}$ & \makecell[l]{$w_0=-0.726$ \\ $w_a=-1.05$} \\ 
        \hline
        JBP  & $w_0 + w_a\frac{z}{(1+z)^2}$ & $(1+z)^{3(1+w_0)}e^{\frac{3}{2}w_a\frac{z^2}{(1+z)^2}}$ & \makecell[l]{$w_0=-0.641$ \\ $w_a=-2.12$} \\ 
        \hline
        BA  & $w_0 + w_a\frac{z(1+z)}{1+z^2}$ & $(1+z)^{3(1+w_0)}(1+z^2)^{\frac{3}{2}w_a}$ & \makecell[l]{ $w_0=-0.770$ \\ $w_a=-0.51$} \\ 
        \hline
        Exp  & $w_0-w_a + w_a e^{\frac{z}{1+z}}$ & $(1+z)^{3(1+w_0-w_a)} \exp\left[3ew_a\int_{\frac{1}{1+z}}^1\frac{e^{-t}}{t}dt\right]$ & \makecell[l]{$w_0=-0.804$ \\ $w_a=-0.71$} \\ 
        \hline
        Log  & $w_0 - w_a\ln\frac{1}{1+z}$ & $(1+z)^{3(1+w_0)}e^{\frac32 w_a\left[\ln(1+z)\right]^2}$ & \makecell[l]{$w_0=-0.763$ \\ $w_a=-0.72$} \\ 
        \hline
        PP  & $\frac{1+z}{3f(z)} \cdot \frac{df(z)}{dz} - 1$ & $1 + f_1\left[1-\frac{1}{1+z}\right] + 
        \frac{f_2}{2}\left[1-\frac{1}{(1+z)^2}\right]$ & \makecell[l]{$f_1=-2.54$ \\ $f_2=3.45$ } \\ 
        \hline
        CG  & $\frac{1+z}{3f(z)} \cdot \frac{df(z)}{dz} - 1$ & $1 + f_1\frac{z}{1+z}  + 
        \frac{f_2}{2}\left(\frac{z}{1+z}\right)^2 + \frac{f_3}{3!}\left(\frac{z}{1+z}\right)^3$ & \makecell[l]{$f_1=0.541$ \\ $f_2=-0.0114$ \\ $f_3=-4.38$} \\ 
        \hline\hline
	\end{tabular}  	
    \end{spacing}	
	\caption{Parameterizations of the equation of state of dynamical dark energy. 
    }
	\label{tab:DE}
\end{table}  

\begin{figure}[t] 
	\begin{center} 
		\includegraphics[width=1\textwidth]{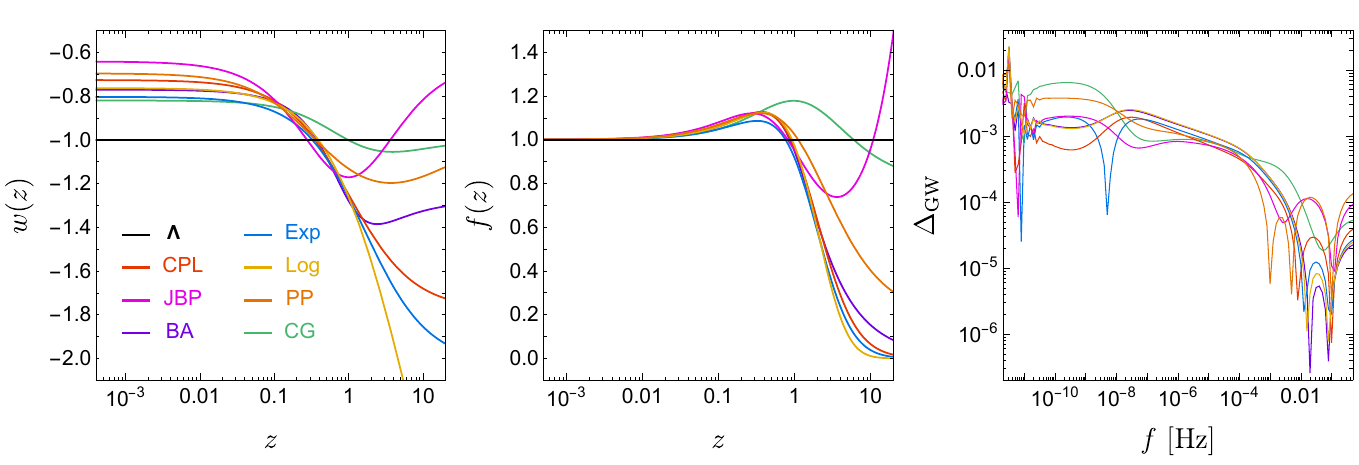}
		\caption{
        The evolution of $w(z)$ (left panel) and $f(z)$ (middle panel) at small redshift for the DDE parameterizations in Table~\ref{tab:DE}. Right panel: The relative deviation $\Delta_{\rm GW}$ in the GW spectra between the DDE models and $\LCDM$.  }
		\label{fig:w_f_GW}
	\end{center}
\end{figure}

Phenomenological parameterizations of $w(z)$ can be divided into two broad classes.
The first class directly assumes a redshift-dependent function for $w(z)$, and includes the Chevallier-Polarski-Linder (CPL) 
parametrization~\cite{Chevallier:2000qy,Linder:2002et}, Jassal-Bagla-Padmanabhan (JBP) parametrization~\cite{Jassal:2005qc}, Barboza-Alcaniz (BA) parametrization~\cite{Barboza:2008rh}, exponential (Exp) parametrization~\cite{Dimakis:2016mip,Pan:2019brc}, and the logarithmic (Log) parametrization~\cite{Efstathiou:1999tm}.
The second class directly parametrizes the dark energy density $\rho_{\rm DE}(z)$, equivalent to parameterizing $f(z)$ and then obtaining $w(z)$ from Eq.~\eqref{eq:fDE}. These include the pressure parametrization (PP)~\cite{Sen:2007gk,Cheng:2025lod} and the cosmographic expansion (CG)~\cite{Visser:2003vq,Capozziello:2011tj,Aviles:2012ay,DES:2024ywx}. Table~\ref{tab:DE} summarizes these parameterizations, and provides representative parameter values for the first class from Ref.~\cite{Giare:2024gpk}, for PP from Ref.~\cite{Cheng:2025lod} and CG from Refs.~\cite{Aviles:2012ay,DES:2024ywx}.
The evolution of $w(z)$ and $f(z)$ with redshift is shown in Fig.~\ref{fig:w_f_GW}.

Using Eqs.~\eqref{eq:P_k},~\eqref{eq:h2Omega_k}, and~\eqref{eq:h2Omegak}, the Hubble expansion rate in Eq.~\eqref{eq:HDDE}, and the loop number density in Eq.~\eqref{eq:n_total_LCDM}, we calculate the GW spectra from cosmic strings in these different late-time cosmological models.
To quantify the difference in the GW spectra between the DDE models and $\LCDM$, we define their relative deviation as
\begin{align}
    \Delta_{\rm GW}(f) = \bigg\vert \frac{ \Omega_{\rm GW}^{\rm DDE}(f) }{\Omega_{\rm GW}^{\Lambda{\rm CDM}}(f)} -1  \bigg\vert\,.
\end{align} 
From the right panel of Fig.~\ref{fig:w_f_GW} it is clear that
 dark energy evolution in the late Universe has a small effect on the GW spectrum of cosmic strings. This is not surprising 
because the GW signal receives contributions from loop emission over a wide range of redshifts. 
Most of the GWs are produced during the RD and MD eras, as shown in Fig.~\ref{fig:GW_spectrum_LCDM}, when the expansion histories of the DDE models are almost identical to that of $\LCDM$.

\section{Summary} \label{sec:6}

A cosmic string network in $\LCDM$ produces a well-defined gravitational wave spectrum which depends only on the string tension. We showed how the GW spectrum is modified if the evolution of the early Universe is altered. 
As a baseline for comparison, we consider a cosmic string network in the scaling regime in the radiation dominated era of $\LCDM$. 
We study four possible modifications to early Universe dynamics by embedding new physics epochs in the radiation era: an era with extra relativistic degrees of freedom (eDOF), an early matter domination era (EMD), an intermediate vacuum domination era (IVD) and a kination era (KIN). A NP era affects the evolution of the string network and consequently modifies loop production. We apply the velocity-dependent one-scale model to account for the NP contribution. Our primary results are:
\begin{itemize}
    \item A NP era can modify the loop number density significantly. The KIN era enhances the loop number density, the EMD and IVD eras suppress it, and the eDOF era leaves it unchanged. Consequently, these eras affect the GW spectrum quite differently. At high frequencies, $\Omega_{\rm GW}^{\rm KIN}>\Omega_{\rm GW}^{\LCDM}\gtrsim\Omega_{\rm GW}^{\rm eDOF}>\Omega_{\rm GW}^{\rm EMD}>\Omega_{\rm GW}^{\rm IVD}$.

    \item For a NP scale $\Lambda_{\rm NP}$, the GW spectrum deviates from that for $\LCDM$ at frequencies above $f \sim (10~{\rm Hz}) \times \Lambda_{\rm NP}/{\rm TeV}$. 
    Thus, modifications in the GW spectrum due to GeV- and TeV-scale NP are detectable at space-based and ground-based GW observatories, respectively. 

    \item For NP eras before BBN, the GW spectrum in the PTA band is unchanged.

    \item Late-time dynamical dark energy has a negligible impact on the GW signal.

\end{itemize}

\section*{Acknowledgements} 
X.H.H and Y.L.Z. are supported by the National Natural Science Foundation of China (NSFC) under Grant Nos.~12535007 and 12547104, and Zhejiang Provincial Natural Science Foundation of China under Grant No. LDQ24A050002.
D.M. is supported in part by the U.S. Department of Energy under Grant No.~de-sc0010504. 

\appendix

\section{Nonscaling effects on the GW spectrum} \label{app:A}

We comment on how nonscaling effects on network evolution affect the GW spectrum in nonstandard cosmology. When the Universe is in a state of constant $\beta \equiv H t$, the network scales, i.e., $\xi = l/t$ is a constant. As the Universe transits between eras, $\beta$ is no longer constant, and $\xi$ varies with time until it approaches a new scaling value. Thus, nonscaling effects appear during the transition. In $\LCDM$, this happens in the RD era (because the number of relativistic DOF varies) and during the transition from the RD to MD era. However, nonscaling effects in 
$\LCDM$ are small~\cite{Marfatia:2023fvh}, as confirmed by Fig.~\ref{fig:GW_spectrum_LCDM}. 

\begin{figure}[t]
	\begin{center} 
		\includegraphics[width=.8\textwidth]{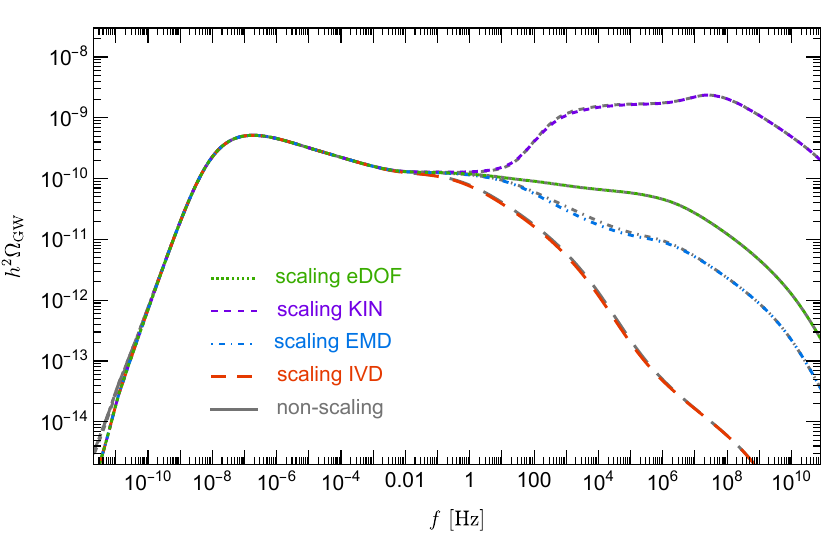}
		\caption{ GW spectra for scaling (colored) and nonscaling (gray) networks for the eDOF (dotted), KIN (short-dashed), EMD (dot-dashed) and IVD (long-dashed) eras between [1~TeV, 10~TeV]. }
		\label{fig:GW_spectrum_scaling}
	\end{center}
\end{figure}

We now consider nonscaling effects in the NP eras.
We apply the nonscaling loop number densities in Eq.~\eqref{eq:n_total}, with Eq.~\eqref{eq:n_r} for $n^{\rm r1}$ and $n^{\rm r2}$, and Eq.~\eqref{eq:n_m} for $n^{\rm m}$.
In the IVD era, $n^{\rm NP}=0$. The loop number densities during the EMD and KIN eras are listed below.
\begin{itemize}

    \item \textbf{EMD.} The loop production function in the EMD era is given by Eq.~\eqref{eq:P_m}, 
    so that the corresponding loop number density is
    \begin{align}
        n^{\rm NP}_{\rm nonscaling}(l,t) = &\ \frac{0.31}{(\alpha_L^{\rm m})^{0.31}} \frac{\mathcal{F}_{\rm m}\tilde{c}}{\gamma_\circ} \frac{1}{t^2} \int_{t_{\rm  NP}^{\rm i}}^t \frac{v_\infty(t')}{[\xi(t')]^{3.31}} \frac{1}{l'^{1.69}} \frac{1}{t'^{1.31}}  \cdot \theta\left(\alpha_L^{\rm m} \xi(t') - \frac{l'}{t'} \right) dt'  \,. \label{eq:n_EMD}
    \end{align}

    \item \textbf{KIN.} Substituting $a(t) \propto t^{1/3}$ into Eq.~\eqref{eq:n_mono}, the number density of loops produced in the KIN era is
    \begin{align}
        n^{\rm NP}_{\rm nonscaling}(l,t) &\,= \frac{{\cal F}_{\rm k}\tilde{c}}{\alpha_L^{\rm k}\gamma_\circ} \cdot \frac{v_\infty^\star}{\xi_\star^4} \, \frac{1}{t^4} \frac{1}{|\alpha_L^{\rm k}\xi_\star+\alpha_L^{\rm k}\dot{\xi}_\star t_\star +\Gamma G\mu|}\,  \left[\frac{\alpha_L^{\rm k} \xi_\star + \Gamma G\mu}{l/t + \Gamma G \mu}\right]^{3}  \,.  \label{eq:n_KIN} 
    \end{align}

\end{itemize}

In Fig.~\ref{fig:GW_spectrum_scaling}, we compare the GW spectra for the scaling and nonscaling solutions and find them to be very slightly different.  
This is consistent Fig.~\ref{fig:xidot}.

\newpage
\bibliographystyle{JHEP}
\bibliography{refs}

\end{document}